\documentclass[12pt]{article}

\usepackage{amsmath,amssymb,amsfonts,amsthm,hyperref,bbm,latexsym,epsfig}
\usepackage{graphicx,multirow,caption}
\usepackage{color}

\newcommand{\bs}{\begin{subequations}}
\newcommand{\es}{\end{subequations}}
\newcommand{\be}{\begin{equation}}
\newcommand{\ee}{\end{equation}}
\newcommand{\ba}{\begin{eqnarray}}
\newcommand{\ea}{\end{eqnarray}}
\newcommand{\no}{\nonumber \\}

\newcommand{\mnu}{\mathcal{M}_\nu}
\newcommand{\nn}{\mathcal{N}}

\allowdisplaybreaks
\begin{document}

\title{
\normalsize \hfill CFTP/19-003
\\[6mm]
\LARGE More models for lepton mixing with four constraints}

\author{
\addtocounter{footnote}{2}
Darius~Jur\v{c}iukonis$^{(1)}$\thanks{\tt darius.jurciukonis@tfai.vu.lt}
\ {\normalsize and}
Lu\'is~Lavoura$^{(2)}$\thanks{\tt balio@cftp.tecnico.ulisboa.pt}
\\*[3mm]
$^{(1)} \! $
\small University of Vilnius,
\small Institute of Theoretical Physics and Astronomy, \\
\small Saul\.{e}tekio ave.~3, LT-10222 Vilnius, Lithuania 
\\[2mm]
$^{(2)} \! $
\small Universidade de Lisboa, Instituto Superior T\'ecnico, CFTP, \\
\small Av.~Rovisco Pais~1, 1049-001 Lisboa, Portugal
\\*[2mm]
}

\date{\today}

\maketitle

\begin{abstract}
  We propose
  new lepton-mixing textures
  that may be enforced through well-defined symmetries
  in renormalizable models.
  Each of our textures has four sum rules for the neutrino mass observables.
  The models are based on the type-I seesaw mechanism;
  their charged-lepton mass matrices are diagonal
  because of the symmetries imposed.
  Each model has three versions,
  depending on the identification of the charged leptons.
  Testing all the models,
  we have found that five of them agree with the data at the $1\sigma$ level
  when the neutrino-mass ordering is normal,
  and two models agree with the data for an inverted ordering.
  We detail the predictions of each of those seven models.
\end{abstract}

\newpage

\section{Introduction and notation}

In this paper we use the type-I seesaw mechanism~\cite{seesaw}
for suppressing the light-neutrino masses.
Let $\ell_L$ and $\ell_R$ be $3 \times 1$ column matrices
that subsume the three left-handed and the three right-handed,
respectively,
charged-lepton fields;
let $\nu_L$ and $\nu_R$ analogously subsume
the three left-handed and the three right-handed neutrino fields.
The lepton mass terms are given by
\be
\mathcal{L}_\mathrm{mass} = - \overline{\ell_L} M_\ell \ell_R
- \overline{\nu_R} M_D \nu_L
- \frac{1}{2}\, \overline{\nu_R} M_R C \overline{\nu_R}^T
+ \mathrm{H.c.},
\ee
where $C$ is the charge-conjugation matrix in Dirac space.
We have added to the Standard Model three right-handed neutrinos
with Majorana mass terms subsumed by the $3 \times 3$ symmetric matrix
(in flavour space)
$M_R$.
In all the models in this paper
the charged-lepton mass matrix $M_\ell$ is diagonal:
\be
\label{mell}
M_\ell = \mathrm{diag} \left( a_e,\, a_\mu,\, a_\tau \right),
\ee
where $\left| a_\alpha \right| = m_\alpha$ for $\alpha = e, \mu, \tau$.
The neutrino Dirac mass matrix $M_D$ is also diagonal in all our models:
\be
M_D = \mathrm{diag} \left( b_e,\, b_\mu,\, b_\tau \right).
\label{md}
\ee
The seesaw mechanism takes place when the matrix $M_R$ is invertible
and its eigenvalues are much larger than the $\left| b_\alpha \right|$.
One then obtains
an effective light-neutrino Majorana mass matrix
\be
\mnu = \mnu^{(1)} + \mnu^{(2)} + \cdots,
\ee
where~\cite{Grimus:2000vj}
\bs
\label{ss}
\ba
\mnu^{(1)} &=& - M_D^T M_R^{-1} M_D, \label{app}
\\*[1mm]
\mnu^{(2)} &=& M_D^T M_R^{-1}\,
\frac{M_D M_D^\dagger {M_R^{-1}}^\ast + {M_R^{-1}}^\ast M_D^\ast M_D^T}{2}\,
M_R^{-1} M_D.
\ea
\es
We shall use the approximation $\mnu = \mnu^{(1)}$.
Therefore,
defining $\nn \equiv \mnu^{-1}$,
one has in our models
\be
\label{nn}
\nn_{\alpha \beta} = - \frac{\left( M_R \right)_{\alpha \beta}}{b_\alpha b_\beta}.
\ee
%
Suppose the $\left| b_\alpha \right|$
are at the Fermi mass scale $m_\mathrm{Fermi}$
and the eigenvalues of $M_R$
are at the much larger mass scale $m_\mathrm{seesaw}$.
Then,
neglecting $\mnu^{(2)}$ as compared to $\mnu^{(1)}$
is an approximation of order
$\left( m_\mathrm{Fermi} \left/ m_\mathrm{seesaw} \right. \right)^2$.
The diagonalization of $\mnu$ proceeds as
\be
\label{diaga}
U^T \mnu U = \mathrm{diag} \left( m_1,\, m_2,\, m_3 \right),
\ee
or
\be
\label{diaga2}
\nn = U \times \mathrm{diag} \left( \frac{1}{m_1},\ \frac{1}{m_2},\
\frac{1}{m_3} \right) \times U^T,
\ee
where $m_{1,2,3}$ are the light-neutrino masses;
they are non-negative real.
Since the charged-lepton mass matrix is diagonal from the start,
$U$ in equation~\eqref{diaga} is the lepton mixing matrix.
We use the parameterization in ref.~\cite{pdg}:
\be
\label{u}
U = \left( \begin{array}{ccc}
  c_{12} c_{13} &
  s_{12} c_{13} e^{i \alpha_{21} / 2} &
  \epsilon^\ast e^{i \alpha_{31} / 2} \\
  - s_{12} c_{23} - c_{12} s_{23} \epsilon &
  \left( c_{12} c_{23} - s_{12} s_{23} \epsilon \right) e^{i \alpha_{21} / 2} &
  s_{23} c_{13} e^{i \alpha_{31} / 2} \\
  s_{12} s_{23} - c_{12} c_{23} \epsilon &
  \left( - c_{12} s_{23} - s_{12} c_{23} \epsilon \right) e^{i \alpha_{21} / 2} &
  c_{23} c_{13} e^{i \alpha_{31} / 2}
\end{array} \right),
\ee
where $\epsilon \equiv s_{13} \exp{\left( i \delta \right)}$,
$c_{ij} = \cos{\theta_{ij}}$,
and $s_{ij} = \sin{\theta_{ij}}$ for $ij = 12,\, 23,\, 13$.
Three different groups
of phenomenologists~\cite{deSalas:2017kay,Capozzi:2018ubv,Esteban:2018azc}
have derived,
from the data provided by various neutrino-oscillation experiments,
values for the mixing angles $\theta_{12,23,13}$,
for the phase $\delta$,
and for the neutrino squared-mass differences.

In general the matrix $\mnu$ determines nine observables:
the three neutrino masses,
the three mixing angles,
the Dirac phase $\delta$,
and the Majorana phases $\alpha_{21}$ and $\alpha_{31}$.
If $\mnu$ contains less than nine independent
rephasing-invariant parameters---\textit{i.e.},
quantities that are invariant under $\left( \mnu \right)_{\alpha \beta}
\to \left( \mnu \right)_{\alpha \beta}
\exp{\left[ i \left( \xi_\alpha + \xi_\beta \right) \right]}$,
where the three phases $\xi_{e,\mu,\tau}$ are arbitrary---then
there will be some relations
(sometimes called `sum rules')
among the nine observables.
This happens in particular when $\mnu$ has two `texture zeroes':
if two out of the six independent matrix elements of $\mnu$ vanish,
then there are four sum rules among the nine observables
(because each vanishing matrix element is in general complex).
Seven viable two-texture-zero cases have been identified
in ref.~\cite{Frampton:2002yf}.\footnote{One of those seven cases (case~C)
  is now excluded by the cosmological upper bound on $m_1 + m_2 + m_3$.}
Other viable cases---or sometimes full models---in which
there are four sum rules among the observables have been discovered,
for instance,
in refs.~\cite{Lavoura:2004tu} and~\cite{Ferreira:2013zqa}.

In this paper we want to present new models with four sum rules that agree,
at the $1\sigma$ level,
with the phenomenological data in at least one of the
three refs.~\cite{deSalas:2017kay,Capozzi:2018ubv,Esteban:2018azc}.
We emphasize that ours are renormalizable models
stabilized by well-defined symmetries;
they are not just ``cases'' or \textit{Ans\"atze}.
We shall present models that predict
\bs
\label{allmodels}
\ba
\label{model1} \label{pred1}
\mathrm{model~1}: & & \nn_{\tau \tau} = 0
\quad \mbox{and}
\quad \nn_{ee} \left( \nn_{\mu \tau} \right)^2
= - \nn_{\mu \mu} \left( \nn_{e \tau} \right)^2,
\\*[1mm]
\label{model5} \label{pred5}
\mathrm{model~2}: & & \nn_{\mu \mu} = 0
\quad \mbox{and} \quad \nn_{ee} \left( \nn_{\mu \tau} \right)^2
= - \nn_{\tau \tau} \left( \nn_{e \mu} \right)^2,
\\*[1mm]
\label{model2} \label{pred2}
\mathrm{model~3}: & & \nn_{e \mu} = 0
\quad \mbox{and} \quad \nn_{ee} \left( \nn_{\mu \tau} \right)^2
= - \nn_{\mu \mu} \left( \nn_{e \tau} \right)^2,
\\*[1mm]
\label{model3} \label{pred3}
\mathrm{model~4}: & & \nn_{\mu \mu} = 0
\quad \mbox{and} \quad \nn_{ee} \left( \nn_{\mu \tau} \right)^2
= \nn_{\tau \tau} \left( \nn_{e \mu} \right)^2,
\\*[1mm]
\label{model4} \label{pred4}
\mathrm{model~5}: & & \nn_{\mu \mu} = 0
\quad \mbox{and} \quad
\left| \nn_{\tau \tau} \left( \nn_{e \mu} \right)^2 \right|^2
- \left| \nn_{ee} \left( \nn_{\mu \tau} \right)^2 \right|^2
\no & &
\quad \quad \quad \quad \quad \quad \quad \ = 2 \left(
\left| \nn_{e \mu} \right|^2 \nn_{\mu \tau} \nn_{e \tau}
\nn_{\tau \tau}^\ast \nn _{e \mu}^\ast
-
\left| \nn_{\mu \tau} \right|^2 \nn_{ee} \nn_{\mu \tau}
\nn_{e \tau}^\ast \nn _{e \mu}^\ast
\right),
\\*[1mm]
\label{model6} \label{pred6}
\mathrm{model~6}: & & \nn_{ee} = 0
\quad \mbox{and} \quad \nn_{\mu \mu} \left( \nn_{e \tau} \right)^2
= \nn_{\tau \tau} \left( \nn_{e \mu} \right)^2,
\\*[1mm]
\label{model7} \label{pred7}
\mathrm{model~7}: & & \nn_{ee} = 0
\quad \mbox{and} \quad
\left| \nn_{\tau \tau} \left( \nn_{e \mu} \right)^2 \right|^2
- \left| \nn_{\mu \mu} \left( \nn_{e \tau} \right)^2 \right|^2
\no & & \quad \quad \quad \quad \quad \quad \quad \
= 2 \left(
\left| \nn_{e \mu} \right|^2 \nn_{\mu \tau} \nn_{e \tau}
\nn_{\tau \tau}^\ast \nn _{e \mu}^\ast
-
\left| \nn_{e \tau} \right|^2 \nn_{\mu \mu} \nn_{e \tau}
\nn_{\mu \tau}^\ast \nn _{e \mu}^\ast
\right). \hspace*{9mm}
\ea
\es
Equations~\eqref{allmodels} may be cast in the simpler form
\bs
\label{allmodels2}
\ba
\mathrm{model~1}: & & \nn_{\tau \tau} = 0
\quad \mbox{and} \quad A_{e \mu} = \frac{1}{2}\,;
\\*[1mm]
\mathrm{model~2}: & & \nn_{\mu \mu} = 0
\quad \mbox{and} \quad A_{e \tau} = \frac{1}{2}\,;
\\*[1mm]
\mathrm{model~3}: & & \nn_{e \mu} = 0
\quad \mbox{and} \quad A_{\tau \tau} = 1\,;
\\*[1mm]
\mathrm{model~4}: & & \nn_{\mu \mu} = 0
\quad \mbox{and} \quad A_{ee} = A_{\tau \tau}\,;
\\*[1mm]
\mathrm{model~5}: & & \nn_{\mu \mu} = 0
\quad \mbox{and} \quad A_{ee} = A_{\tau \tau}^\ast\,;
\label{model52} \\*[1mm]
\mathrm{model~6}: & & \nn_{ee} = 0
\quad \mbox{and} \quad A_{\mu \mu} = A_{\tau \tau}\,;
\\*[1mm]
\mathrm{model~7}: & & \nn_{ee} = 0
\quad \mbox{and} \quad A_{\mu \mu} = A_{\tau \tau}^\ast,
\ea
\es
where the matrix $A$ is defined through~\cite{Ferreira:2013oga}
\be
\label{a1}
A_{\alpha\beta} \equiv \nn_{\alpha\beta} \left( \nn^{-1} \right)_{\alpha \beta}
= \left( \mnu \right)_{\alpha \beta} \left( \mnu^{-1} \right)_{\alpha \beta}
\ee
(no summation over $\alpha$ and $\beta$ is understood).
In our models,
because of equation~\eqref{nn},
\be
A_{\alpha \beta} =
\left( M_R \right)_{\alpha \beta} \left( M_R^{-1} \right)_{\alpha \beta}.
\ee

In section~\ref{sec2} we shall present models~1 and~3.
In section~\ref{sec3} we shall present models~4 and~5.
Since equations~\eqref{pred5} are the same as equations~\eqref{pred1}
after a $\mu$--$\tau$ interchange,
and since equations~\eqref{pred6} and~\eqref{pred7}
are the same as equations~\eqref{pred3} and~\eqref{pred4},
respectively,
after an $e$--$\mu$ interchange,
our models~1,
4,
and~5 can also be identified as models~2,
6,
and~7,
respectively,
if one labels the charged leptons in a different manner.
An analysis of the practical consequences of our sum rules
is deferred to section~\ref{sec4};
it turns out that models~1--5
agree with the data at the 1$\sigma$ level when the neutrino mass ordering
is normal (`NO'),
\textit{viz.}\ $m_1 < m_2 < m_3$,
while models~6 and~7
agree with the data at the 1$\sigma$ level when the neutrino mass ordering
is inverted (`IO'),
\textit{viz.}\ $m_3 < m_1 < m_2$.
A short summary of our findings is attempted in section~\ref{sec5}.

\section{Models~1 and~3}
\label{sec2}

The models in this section are inspired by those
in ref.~\cite{Grimus:2005jk},
\textit{viz.}\ they are based on the idea of a
(leading-order)
antisymmetry of $M_R$ under an $e$--$\mu$ interchange.

All the models in this paper have gauge group $SU(2) \times U(1)$.
There are three left-handed-lepton gauge-$SU(2)$ doublets
$D_\alpha = \left( \nu_{\alpha L},\ \alpha_L \right)^T$,
three right-handed charged-lepton $SU(2)$ singlets $\alpha_R$,
and three right-handed-neutrino gauge singlets $\nu_{\alpha R}$.
In all the models in this paper
we use two scalar gauge-$SU(2)$ doublets $\phi_1$
and $\phi_2$.\footnote{The models in ref.~\cite{Grimus:2005jk}
  had three scalar doublets.
  In the models of this paper we need only two.}
Let $v_a$ ($a = 1, 2$) denote
the vacuum expectation values (VEVs) of the neutral components $\phi_a^0$
of $\phi_a = \left( \phi_a^+,\ \phi_a^0 \right)^T$.
We define $\tilde \phi_a \equiv i \tau_2 \phi_a^\ast
= \left( {\phi_a^0}^\ast,\ - \phi_a^- \right)$.

In the models in this section there is
\emph{one complex scalar gauge singlet $S$}.
We introduce the flavour-lepton-number symmetries $L_\alpha$;
the dimension-four terms in the Lagrangian respect those symmetries
but lower-dimension terms are allowed to break them.
Thus,
in these models there is \emph{soft}\/ symmetry breaking
(besides spontaneous symmetry breaking).\footnote{Soft (super)symmetry breaking
  is widely used in model-buiding---notably,
  it is always used in supersymmetric model building.
  Soft breaking consists in a symmetry holding
  in all the Lagrangian terms of dimension higher than some value,
  but not holding for the Lagrangian terms of dimension smaller than,
  or equal to,
  that value.
  In our case,
  the family-lepton-number symmetries hold for terms of dimension four
  but are broken by terms of dimension three,
  \textit{viz.}\ the terms in equation~\eqref{gmkfp}.
  In principle,
  a model with a softly broken symmetry should eventually be justified
  through an ultraviolet completion,
  \textit{viz.}\ a more complete model,
  with extra fields active at higher energies,
  which effectively mimics at lower energy scales
  the model with the softly-broken symmetry.
  Unfortunately,
  an ultraviolet completion may be difficult to construct explicitely.
  In its absence,
  a softly broken (super)symmetry constitutes a non-trivial assumption.
  This may be considered to be a weakness of models~1--3 in this paper.}
The multiplets $D_\alpha$,
$\alpha_R$,
and $\nu_{\alpha R}$ have $U(1)$ charge $+1$ under $L_\alpha$
and $U(1)$ charges $0$ under the $L_\beta$ with $\beta \neq \alpha$.
We also enforce a $\mathbbm{Z}_4$ symmetry
that interchanges $e$ and $\mu$\footnote{The full symmetry group
  of the model is $G = \left\{ \left[ U(1)_{L_e} \times U(1)_{L_\mu} \right]
  \rtimes \mathbbm{Z}_4 \right\} \times U(1)_{L_\tau}$.
  Here,
  the $\mathbbm{Z}_4$ subgroup of $G$ is formed by the matrices
  \be
  \left( \begin{array}{ccc} 1 & 0 & 0 \\
    0 & 1 & 0 \\ 0 & 0 & 1 \end{array} \right), \quad
  \left( \begin{array}{ccc} 0 & i & 0 \\
    i & 0 & 0 \\ 0 & 0 & 1 \end{array} \right), \quad
  \left( \begin{array}{ccc} -1 & 0 & 0 \\
    0 & -1 & 0 \\ 0 & 0 & 1 \end{array} \right), \quad
  \left( \begin{array}{ccc} 0 & -i & 0 \\
    -i & 0 & 0 \\ 0 & 0 & 1 \end{array} \right).
  \ee
  The normal subgroup
  $N = U(1)_{L_e} \times U(1)_{L_\mu} \times U(1)_{L_\tau}$ of $G$
  is formed by the matrices
  \be
  \left( \begin{array}{ccc} e^{i \left( p \alpha + q \beta \right)} & 0 & 0 \\
    0 & e^{i \left( r \alpha + s \beta \right)} & 0 \\
    0 & 0 & e^{i t \gamma} \end{array} \right),
  \ee
  where $p$, $q$, $r$, $s$, and $t$ are integers and $\alpha$,
  $\beta$, and $\gamma$ are the phases that generate $U(1)_{L_e}$,
  $U(1)_{L_\mu}$,
  and $U(1)_{L_\tau}$,
  respectively.
  Every matrix $g \in G$ may be written in a unique way as $g = n h$,
  where $n \in N$ and $h \in \mathbbm{Z}_4$.
  The multiplication rule is $\left( n h,\ n^\prime h^\prime \right) =
  \left( n h n^\prime h^{-1},\ h h^\prime \right)$;
  notice that $h n^\prime h^{-1} \in N$ because $N$ is a normal subgroup,
  hence $n h n^\prime h^{-1}$ is also in $N$.}:
\bs
\label{zzzz4444}
\ba
& & D_e \to i D_\mu, \quad D_\mu \to i D_e, \quad D_\tau \to i D_\tau, \\
& & e_R \to i \mu_R, \quad \mu_R \to i e_R, \quad \tau_R \to i \tau_R, \\
& & \nu_{eR} \to i \nu_{\mu R}, \quad \nu_{\mu R} \to i \nu_{eR}, \quad
\nu_{\tau R} \to i \nu_{\tau R}, \\
& & \phi_2 \to - \phi_2, \quad S \to - S.
\ea
\es
The Yukawa Lagrangian coupling the leptons to the scalar doublets is therefore
\bs
\ba
\mathcal{L}_{\mathrm{Y} \phi} &=&
- y_1 \overline{D_\tau} \tau_R \phi_1
- y_2 \left( \overline{D_e} e_R + \overline{D_\mu} \mu_R \right) \phi_1
\\ & &
- y_3 \left( \overline{D_e} e_R - \overline{D_\mu} \mu_R \right) \phi_2
\label{l2} \\ & &
- y_4 \overline{D_\tau} \nu_{\tau R} \tilde \phi_1
- y_5 \left( \overline{D_e} \nu_{eR}
+ \overline{D_\mu} \nu_{\mu R} \right) \tilde \phi_1
\\ & &
- y_6 \left( \overline{D_e} \nu_{eR}
- \overline{D_\mu} \nu_{\mu R} \right) \tilde \phi_2
+ \mathrm{H.c.}
\label{l4}
\ea
\es
Therefore,
the charged-lepton mass matrix and the neutrino Dirac mass matrix are
diagonal as anticipated in equations~\eqref{mell} and~\eqref{md},
respectively,
with\footnote{Since $\left| a_e \right| = m_e \ll \left| a_\mu \right| = m_\mu$,
  a finetuning is necessary to make $y_3 v_2 \approx - y_2 v_1$.
  This finetuning may be justified through
  an additional symmetry~\cite{Grimus:2003vx}.
  We shall not pursue that idea here.}
\be
\label{ab}
\begin{array}{rclcrclcrcl}
a_\tau &=& y_1 v_1,  & & a_e &=& y_2 v_1 + y_3 v_2, & &
a_\mu &=& y_2 v_1 - y_3 v_2,
\\*[1mm]
b_\tau &=& y_4^\ast v_1, & & b_e &=& y_5^\ast v_1 + y_6^\ast v_2, & &
b_\mu &=& y_5^\ast v_1 - y_6^\ast v_2.
\end{array}
\ee
The doublet $\phi_2$ and its Yukawa couplings in lines~\eqref{l2}
and~\eqref{l4} are needed so that $m_e \neq m_\mu$ and $b_e \neq b_\mu$.

There are right-handed-neutrino Majorana mass terms
\be
\label{gmkfp}
\mathcal{L}_{\mathrm{M} \nu} = \frac{m^\ast}{2} \left(
\nu_{eR}^T C^{-1} \nu_{eR} - \nu_{\mu R}^T C^{-1} \nu_{\mu R} \right)
+ {m^\prime}^\ast \nu_{\tau R}^T C^{-1} \left(
\nu_{eR} - \nu_{\mu R} \right)
+ \mathrm{H.c.}
\ee
The terms in $\mathcal{L}_{\mathrm{M} \nu}$
violate the family-lepton-number symmetries $L_\alpha$;
this is allowed because those terms have mass dimension three.
However,
$\mathcal{L}_{\mathrm{M} \nu}$ is not allowed to break $\mathbbm{Z}_4$,
which is broken spontaneously but not softly.

\paragraph{Model~1:} In this model
the singlet $S$ has $L_e = L_\mu = +1$ and $L_\tau = 0$.
There is then a coupling
\be
\mathcal{L}_S = y_s S\, \overline{\nu_{eR}}\, C\, \overline{\nu_{\mu R}}^T
+ \mathrm{H.c.},
\ee
where $y_s$ is a Yukawa coupling constant.
The Majorana mass matrix of the right-handed neutrinos is
\be
M_R = \left( \begin{array}{ccc}
  m & y_s w & m^\prime \\ y_s w & -m & -m^\prime \\ m^\prime & -m^\prime & 0
\end{array} \right),
\ee
%
where $w$ is the VEV of $S$.\footnote{We assume that $y_s w$,
  $m$,
  and $m^\prime$ are all of the same order of magnitude $m_\mathrm{seesaw}$.}
Using equation~\eqref{nn},
it is now obvious that equations~\eqref{pred1} hold.

Because the family-lepton-number symmetries are softly broken,
terms proportional to $S^2$,
$\phi_1^\dagger \phi_2 S$,
and $\phi_1^\dagger \phi_2 S^\ast$
(and their Hermitian conjugates)
are present in the scalar potential
even while $S$ carries family lepton numbers.
Those terms eliminate the Goldstone boson that would appear
if the (continuous) family-lepton-number symmetries
were broken solely through $w \neq 0$.

\paragraph{Model~3:} In this model
the singlet $S$ has $L_e = L_\mu = 0$ and $L_\tau = +2$.
There is a coupling
\be
\mathcal{L}_S = \frac{y_s S}{2}\,
\overline{\nu_{\tau R}}\, C\, \overline{\nu_{\tau R}}^T
+ \mathrm{H.c.}
\ee
Then,
\be
M_R = \left( \begin{array}{ccc}
  m & 0 & m^\prime \\ 0 & -m & -m^\prime \\ m^\prime & -m^\prime & y_s w
\end{array} \right).
\ee
The matrix $\nn$ then satisfies equations~\eqref{pred2}.

\section{Models~4 and~5}
\label{sec3}

The models in this section use
\emph{two complex scalar gauge singlets $S_1$ and $S_2$
  and one real singlet $S_3$};
thus,
their scalar sector is larger than the one
of the models of the previous section.
In this section \emph{we do not employ soft symmetry breaking}.
We use a symmetry $\mathbbm{Z}_4^{(1)} \times \mathbbm{Z}_4^{(2)}$,
where
\bs
\label{zzzzzzzz}
\ba
\mathbbm{Z}_4^{(1)}: & & \left\{ \begin{array}{l}
\left( e_R,\ \nu_{eR},\ D_e \right) \to i \left( e_R,\ \nu_{eR},\ D_e \right),
\\*[1mm]
\left( \tau_R,\ \nu_{\tau R},\ D_\tau \right)
\to - i \left( \tau_R,\ \nu_{\tau R},\ D_\tau \right),
\\*[1mm]
S_1 \to i S_1,\  S_2 \to - i S_2,\ S_3 \to - S_3,
\end{array} \right.
\label{zzzzzzzz1}
\\*[2mm]
\mathbbm{Z}_4^{(2)}: & & \left\{ \begin{array}{l}
  \left( \mu_R,\ \nu_{\mu R},\ D_\mu \right)
  \to i \left( \mu_R,\ \nu_{\mu R},\ D_\mu \right),
\\*[1mm]
S_1 \to i S_1,\ S_2 \to i S_2.
\end{array} \right.
\label{zzzzzzzz2}
\ea
\es
This symmetry allows for the Yukawa Lagrangian
\bs
\label{djkdpr}
\ba
\mathcal{L}_\mathrm{Y} &=&
- \left( y_1 \overline{D_\mu} \mu_R
+ y_2 \overline{D_e} e_R
+ y_3 \overline{D_\tau} \tau_R \right) \phi_1
\\ & &
- \left( y_4 \overline{D_\mu} \mu_R
+ y_5 \overline{D_e} e_R
+ y_6 \overline{D_\tau} \tau_R \right) \phi_2
\\ & &
- \left( y_7 \overline{D_\mu} \nu_{\mu R}
+ y_8 \overline{D_e} \nu_{eR}
+ y_9 \overline{D_\tau} \nu_{\tau R} \right) \tilde \phi_1
\\ & &
- \left( y_{10} \overline{D_\mu} \nu_{\mu R}
+ y_{11} \overline{D_e} \nu_{eR}
+ y_{12} \overline{D_\tau} \nu_{\tau R} \right) \tilde \phi_2
\\ & &
- \overline{\nu_{\mu R}} C
\left( y_{13} \overline{\nu_{eR}}^T S_1
+ y_{14} \overline{\nu_{\tau R}}^T S_2 \right)
\\ & &
- \frac{y_{15}}{2}\, \overline{\nu_{eR}} C \overline{\nu_{eR}}^T S_3
- \frac{y_{16}}{2}\, \overline{\nu_{\tau R}} C \overline{\nu_{\tau R}}^T S_3
+ \mathrm{H.c.}
\ea
\es
The charged-lepton mass matrix and the neutrino Dirac mass matrix
are given by equations~\eqref{mell} and~\eqref{md},
respectively,
with
\be
\label{ab2}
\begin{array}{rclcrclcrcl}
  a_\mu &=& y_1 v_1 + y_4 v_2, & &
  a_e &=& y_2 v_1 + y_5 v_2, & &
  a_\tau &=& y_3 v_1 + y_6 v_2,
  \\*[1mm]
  b_\mu &=& y_7^\ast v_1 + y_{10}^\ast v_2, & &
  b_e &=& y_8^\ast v_1 + y_{11}^\ast v_2, & &
  b_\tau &=& y_9^\ast v_1 + y_{12}^\ast v_2.
\end{array}
\ee
The symmetry~\eqref{zzzzzzzz} also allows a bare Majorana mass term
\be
\label{bdjiho}
- m\, \overline{\nu_{eR}} C \overline{\nu_{\tau R}}^T + \mathrm{H.c.}
\ee
The Majorana mass matrix of the right-handed neutrinos is then
\be
\label{mr}
M_R = \left( \begin{array}{ccc}
  y_{15} w_3 & y_{13} w_1 & m \\
  y_{13} w_1 & 0 & y_{14} w_2 \\ 
  m & y_{14} w_2 & y_{16} w_3
\end{array} \right),
\ee
where $w_k = \left\langle 0 \left| S_k \right| 0 \right\rangle$
for $k = 1, 2, 3$.
Note that $w_3$ is real because $S_3$ is a real scalar field.
The matrix element $\left( M_R \right)_{22}$ is zero
because of the symmetry $\mathbbm{Z}_4^{(2)}$.
We assume $\left| y_{13} w_1 \right|$,
$\left| y_{14} w_2 \right|$,
$\left| y_{15} w_3 \right|$,
$\left| y_{16} w_3 \right|$,
and $m$ to be all at the same order of magnitude $m_\mathrm{seesaw}$.

\subsection{Model~4}

In model~4 there is an additional $\mathbbm{Z}_2$ symmetry
\be
\label{z2}
\mathbbm{Z}_2: \quad
e_R \leftrightarrow \tau_R,\ \nu_{eR} \leftrightarrow \nu_{\tau R},\
D_e \leftrightarrow D_\tau,\ S_1 \leftrightarrow S_2,\ \phi_2 \to - \phi_2.
\ee
This symmetry does not constrain the bare mass term~\eqref{bdjiho};
in the Yukawa Lagrangian~\eqref{djkdpr} it makes
\bs
\ba
& & y_3 = y_2, \quad y_4 = 0, \quad y_6 = - y_5, \\
& & y_9 = y_8, \quad y_{10} = 0, \quad y_{12} = - y_{11}, \\
& & y_{14} = y_{13}, \quad y_{16} = y_{15},
\label{c}
\ea
\es
so that
\be
m_\mu = \left| y_1 v_1 \right|, \quad
m_e = \left| y_2 v_1 + y_5 v_2 \right|, \quad
m_\tau = \left| y_2 v_1 - y_5 v_2 \right|
\ee
recquires a finetuning to make $y_5 v_2 \approx - y_2 v_1$. 
Because of equations~\eqref{c},
we now have
\be
M_R = \left( \begin{array}{ccc}
  y_{15} w_3 & y_{13} w_1 & m \\
  y_{13} w_1 & 0 & y_{13} w_2 \\ 
  m & y_{13} w_2 & y_{15} w_3
\end{array} \right)
\ee
instead of equation~\eqref{mr}.
Then assuming $w_1^2 = w_2^2$,
one recovers equation~\eqref{pred3} as desired.

The symmetries $\mathbbm{Z}_4^{(1)}$ of equation~\eqref{zzzzzzzz1}
and $\mathbbm{Z}_2$ of equation~\eqref{z2} together
generate the non-Abelian group $D_8$
(the dihedral group with eight elements).
The symmetry $\mathbbm{Z}_4^{(2)}$ of equation~\eqref{zzzzzzzz2} commutes
with both $\mathbbm{Z}_4^{(1)}$ and $\mathbbm{Z}_2$,
\textit{i.e.}\ it commutes with $D_8$.
The group $D_8$ has five irreducible representations:
the $\mathbf{2}$ and the $\mathbf{1}_{pq}$,
where both $p$ and $q$ may be either $+1$ or $-1$.
The Clebsch--Gordan series are
\be
\mathbf{2} \otimes \mathbf{2} = \mathbf{1}_{++} \oplus
\mathbf{1}_{--} \oplus \mathbf{1}_{+-} \oplus \mathbf{1}_{-+},
\quad
\mathbf{2} \otimes \mathbf{1}_{pq} = \mathbf{2},
\quad
\mathbf{1}_{pq} \otimes \mathbf{1}_{p^\prime q^\prime} =
\mathbf{1}_{p p^\prime, q q^\prime}.
\ee
Under $D_8$,
\bs
\ba
\mu_R,\ \nu_{\mu R},\ D_\mu,\ \phi_1 &\mbox{are}& \mathbf{1}_{++},
\\
\phi_2 &\mbox{is}& \mathbf{1}_{+-},
\\
S_3 &\mbox{is}& \mathbf{1}_{-+},
\\
\left( \begin{array}{c} e_R \\ \tau_R \end{array} \right),\
\left( \begin{array}{c} \nu_{eR} \\ \nu_{\tau R} \end{array} \right),\
\left( \begin{array}{c} D_e \\ D_\tau \end{array} \right),\
\left( \begin{array}{c} S_1 \\ S_2 \end{array} \right)
&\mbox{are}& \mathbf{2}.
\ea
\es

In order to justify the assumption $w_1^2 = w_2^2$,
one must look at the potential of the scalar singlets,
which is
\bs
\label{13}
\ba
V_S &=&
\mu_1 \left( \left| S_1 \right|^2 + \left| S_2 \right|^2 \right)
+ \mu_2 S_3^2
+ \lambda_1 \left( \left| S_1 \right|^2 + \left| S_2 \right|^2 \right)^2
+ \lambda_2 S_3^4
+ \lambda_3 \left( \left| S_1 \right|^2 + \left| S_2 \right|^2 \right) S_3^2
\hspace*{10mm}
\\ & &
+ 4 \lambda_4 \left| S_1 S_2 \right|^2
+ \bar m S_3 \left( S_1^\ast S_2 + S_2^\ast S_1 \right)
+ 2 \lambda_5 \left[ \left( S_1^\ast S_2 \right)^2
  + \left( S_2^\ast S_1 \right)^2 \right]
\\ & &
+ \left[ \lambda_6 \left( S_1^4 + S_2^4 \right)
  + 2 \lambda_7 \left( S_1 S_2 \right)^2 + \mathrm{H.c.} \right],
\ea
\es
where $\lambda_6$ and $\lambda_7$ are complex
and all the other couplings are real.
We write
\bs
\ba
w_1 &=& w\, \cos{\frac{\theta}{2}}\ e^{i \chi / 4}\, e^{- i \psi / 2},
\\
w_2 &=& w\, \sin{\frac{\theta}{2}}\ e^{i \chi / 4}\, e^{i \psi / 2},
\ea
\es
where $w$ is positive and $0 \le \theta \le \pi$.
Defining $V_0 \equiv \left\langle 0 \left| V_S \right| 0 \right\rangle$,
we then have
\bs
\ba
V_0 &=&
\mu_1 w^2 + \mu_2 w_3^2 + \lambda_1 w^4 + \lambda_2 w_3^4 + \lambda_3 w^2 w_3^2
\label{14a} \\ & &
+ \lambda_4 w^4 \sin^2{\theta}
+ \bar m w_3 w^2 \sin{\theta} \cos{\psi}
+ \lambda_5 w^4 \sin^2{\theta} \cos{\left( 2 \psi \right)}
\label{14b} \\ & &
+ w^4 \cos{\chi} \left[
  \left( 1 + \cos^2{\theta} \right) \cos{\left( 2 \psi \right)}\,
  \Re{\lambda_6}
  + 2 \cos{\theta} \sin{\left( 2 \psi \right)}\, \Im{\lambda_6}
  + \sin^2{\theta}\, \Re{\lambda_7} \right]
\\ & &
+ w^4 \sin{\chi} \left[
  2 \cos{\theta} \sin{\left( 2 \psi \right)}\, \Re{\lambda_6}
  - \left( 1 + \cos^2{\theta} \right) \cos{\left( 2 \psi \right)}\,
  \Im{\lambda_6}
  - \sin^2{\theta}\, \Im{\lambda_7} \right].
\hspace*{9mm}
\ea
\es
This may be minimized relative to the phase $\chi$,
yielding
\bs
\label{function}
\ba
V_0 &=&
\mbox{lines}\ \eqref{14a}\ \mbox{and}\ \eqref{14b}
\\ & &
- l w^4 \left\{
\left[ 4 \cos^2{\theta} + \sin^4{\theta} \cos^2{\left( 2 \psi \right)} \right]
\cos^2{\alpha} + \sin^4{\theta} \sin^2{\alpha}
\right. \\ & &
+ \left( 1 + \cos^2{\theta} \right) \sin^2{\theta} \cos{\left( 2 \psi \right)}
\sin{\left( 2 \alpha \right)} \cos{\lambda}
\\ & & \left.
+ 2 \cos{\theta} \sin^2{\theta} \sin{\left( 2 \psi \right)}
\sin{\left( 2 \alpha \right)} \sin{\lambda}
\right\}^{1/2},
\ea
\es
where the square root is non-negative and we have defined $l > 0$,
$\alpha \in \left[ 0,\ \pi / 2 \right]$,
and the phase $\lambda$ through
\bs
\ba
\left| \lambda_6 \right| &=& l \cos{\alpha}, \\
\left| \lambda_7 \right| &=& l \sin{\alpha}, \\
2 \lambda_6 \lambda_7^\ast &=& l^2 \sin{\left( 2 \alpha \right)}\, e^{i \lambda}.
\ea
\es
In order to justify $w_1^2 = w_2^2$,
we want the minimum of $V_0$ in equation~\eqref{function} to lie
either at $\theta = \pi / 2$ or $\theta = 3 \pi / 2$,
together with either $\psi = 0$ or $\psi = \pi$,
without necessitating the parameters of the potential
to obey any constraining equation.
We have therefore looked for the minimum of the function
\bs
\ba
f \left( \theta, \psi \right) &=& A \sin^2{\theta}
+ B \sin{\theta} \cos{\psi}
+ C \sin^2{\theta} \cos{\left( 2 \psi \right)}
\\ & &
- \left\{ \left[ 4 \cos^2{\theta}
  + \sin^4{\theta} \cos^2{\left( 2 \psi \right)} \right] \cos^2{\alpha}
+ \sin^4{\theta} \sin^2{\alpha}
\right. \\ & &
+ \left( 1 + \cos^2{\theta} \right) \sin^2{\theta} \cos{\left( 2 \psi \right)}
\sin{\left( 2 \alpha \right)} \cos{\lambda}
\\ & & \left.
- 2 \cos{\theta} \sin^2{\theta} \sin{\left( 2 \psi \right)}
\sin{\left( 2 \alpha \right)} \sin{\lambda} \right\}^{1/2}
\ea
\es
for various values of the parameters $A \equiv \lambda_4 / l$,
$B \equiv \bar m w_3 / l$,
$C \equiv \lambda_5 / l$,
$\alpha \equiv \arctan{\left| \lambda_7 / \lambda_6 \right|}$,
and $\lambda \equiv \arg{\left( \lambda_6 \lambda_7^\ast \right)}$.
We have discovered that,
for instance in the continuous domain
$-0.25 < A < -0.10$,
$-7 < B < -3$,
$-6 < C < -2$,
$0.2 < \alpha < 0.6$,
and $\pi < \lambda < 2 \pi$ the minimum of $f \left( \theta, \psi \right)$
always lies at the desired point $\theta = \pi / 2, \psi = 0$.
Thus,
$w_1 = w_2$ is a possible absolute minimum of the potential
and does not require its parameters to obey any constraint equation.

At low energy,
the minimum with $w_1^2 = w_2^2$
will be perturbed by the term in the potential
$\phi_1^\dagger \phi_2
\left( \left| S_1 \right|^2 - \left| S_2 \right|^2 \right)$,
which is invariant under $D_8 \times \mathbbm{Z}_4^{(2)}$.
However,
that term is of order $\left( m_\mathrm{Fermi}
\left/ m_\mathrm{seesaw} \right. \right)^2 \ll 1$
relative to the potential in equation~\eqref{13}.
We neglect that term just as we neglect $\mnu^{(2)}$
when compared to $\mnu^{(1)}$ in equation~\eqref{ss};
we consistently work in the approximation
$\left( m_\mathrm{Fermi} \left/ m_\mathrm{seesaw} \right. \right)^2 \to 0$.

Terms in the scalar potential
like $\phi_1^\dagger \phi_1 \left( S_3 \right)^2$
and $\phi_1^\dagger \phi_2
\left( \left| S_1 \right|^2 - \left| S_2 \right|^2 \right)^2$
tend to draw $m_\mathrm{Fermi}$
(the mass scale of the VEVs $v_1$ and $v_2$)
to the vicinity of $m_\mathrm{seesaw}$
(the mass scale of $w$ and $w_3$).
This problem arises in any model with two very distinct mass scales
in the scalar sector;
we have no cure to offer to it.

\subsection{Model~5}

Instead of the $\mathbbm{Z}_2$ symmetry~\eqref{z2},
in model~5 we employ the $CP$ symmetry\footnote{The $CP$ symmetry
  must be extended to the quark sector.
  It must be spontaneously broken,
  since we know that $CP$ is not a symmetry of Nature.
  The detailed treatment of those important issues
  is beyond the scope of this paper.}
%
\be
\label{cp}
\begin{array}{lclcl}
\mu_R \left( x \right) \to
i \gamma_0 C \overline{\mu_R}^T \left( \bar x \right),
& \ &
e_R \left( x \right) \to
i \gamma_0 C \overline{\tau_R}^T \left( \bar x \right),
& \ &
\tau_R \left( x \right) \to
i \gamma_0 C \overline{e_R}^T \left( \bar x \right),
\\*[1mm]
\nu_{\mu R} \left( x \right) \to
i \gamma_0 C \overline{\nu_{\mu R}}^T \left( \bar x \right),
& \ &
\nu_{eR} \left( x \right) \to
i \gamma_0 C \overline{\nu_{\tau R}}^T \left( \bar x \right),
& \ &
\nu_{\tau R} \left( x \right) \to
i \gamma_0 C \overline{\nu_{eR}}^T \left( \bar x \right),
\\*[1mm]
D_\mu \left( x \right) \to
i \gamma_0 C \overline{D_\mu}^T \left( \bar x \right),
& \ &
D_e \left( x \right) \to
i \gamma_0 C \overline{D_\tau}^T \left( \bar x \right),
& \ &
D_\tau \left( x \right) \to i \gamma_0 C
\overline{D_e}^T \left( \bar x \right),
\\*[1mm]
\phi_1 \left( x \right) \to \phi_1^\ast \left( \bar x \right),
& \ &
\phi_2 \left( x \right) \to - \phi_2^\ast \left( \bar x \right),
& \ &
\\*[1mm]
S_1 \left( x \right) \leftrightarrow S_2^\ast \left( \bar x \right),
& \ &
S_3 \left( x \right) \to S_3 \left( \bar x \right),
& \ &
\end{array}
\ee
where $x = \left( t,\, \vec r \right)$
and $\bar x = \left( t,\, - \vec r \right)$.
This $CP$ symmetry is compatible with the symmetries
$\mathbbm{Z}_4^{(1)}$ and $\mathbbm{Z}_4^{(2)}$ in equations~\eqref{zzzzzzzz}.
Indeed,
it may easily be verified that the $CP$ transformation~\eqref{cp}
followed by the $\mathbbm{Z}_4^{(1)}$ transformation
and followed by the transformation $\left( CP \right)^{-1}$
is identical to $\mathbbm{Z}_4^{(1)}$;
while the successive application of $CP$,
$\mathbbm{Z}_4^{(2)}$,
and $\left( CP \right)^{-1}$ is equivalent to the successive application
of $\mathbbm{Z}_4^{(2)}$ three times.\footnote{Note that
  $\left( CP \right)^{-1} = CP$ for bosons
  but $\left( CP \right)^{-1} = - CP$ for fermions.}
This demonstrates the compatibility~\cite{ziegler}.

In the Lagrangian~\eqref{djkdpr},
the $CP$ symmetry~\eqref{cp} enforces
\bs
\ba
& & y_1^\ast = y_1, \quad y_3^\ast = y_2, \quad
y_4^\ast = - y_4, \quad y_6^\ast = - y_5, \\
& & y_7^\ast = y_7, \quad y_9^\ast = y_8, \quad
y_{10}^\ast = - y_{10}, \quad y_{12}^\ast = - y_{11}, \\
& & y_{14}^\ast = y_{13}, \quad y_{16}^\ast = y_{15}, \label{uviidpd}
\ea
\es
hence
\be
m_\mu = \left| y_1 v_1 + y_4 v_2 \right|, \quad
m_e = \left| y_2 v_1 + y_5 v_2 \right|, \quad
m_\tau = \left| y_2^\ast v_1 - y_5^\ast v_2 \right|.
\ee
Moreover,
in equation~\eqref{bdjiho} $m$ becomes real.
Because of equation~\eqref{uviidpd} one has
\be
M_R = \left( \begin{array}{ccc}
  y_{15} w_3 & y_{13} w_1 & m \\
  y_{13} w_1 & 0 & y_{13}^\ast w_2 \\ 
  m & y_{13}^\ast w_2 & y_{15}^\ast w_3
\end{array} \right),
\ee
with real $m$.
If one assumes $\left| w_1 \right| = \left| w_2 \right|$,\footnote{One
  does \emph{not}\/ need to assume $w_1 = w_2^\ast$;
  indeed,
  $\left| w_1 \right| = \left| w_2 \right|$ suffices.}
then one recovers equations~\eqref{model52}.

With the $CP$ symmetry~\eqref{cp}
instead of the $\mathbbm{Z}_2$ symmetry~\eqref{z2},
the potential of the scalar singlets is
\bs
\label{132}
\ba
V_S &=&
\mu_1 \left( \left| S_1 \right|^2 + \left| S_2 \right|^2 \right)
+ \mu_2 S_3^2
+ \lambda_1 \left( \left| S_1 \right|^2 + \left| S_2 \right|^2 \right)^2
+ \lambda_2 S_3^4
+ \lambda_3 \left( \left| S_1 \right|^2 + \left| S_2 \right|^2 \right) S_3^2
\hspace*{10mm}
\\ & &
+ 4 \lambda_4 \left| S_1 S_2 \right|^2
+ \left[ \bar m S_3 S_1^\ast S_2 + 2 \lambda_5 \left( S_1^\ast S_2 \right)^2
  + \lambda_6 \left( S_1^4 + {S_2^\ast}^4 \right) + \mathrm{H.c.} \right]
\\ & &
+ 2 \lambda_7 \left[ \left( S_1 S_2 \right)^2 + \mathrm{H.c.} \right],
\ea
\es
with complex $\bar m$,
$\lambda_5$,
and $\lambda_6$ but real $\lambda_7$.
Hence,
\bs
\ba
V_0 &=&
\mu_1 w^2 + \mu_2 w_3^2 + \lambda_1 w^4 + \lambda_2 w_3^4 + \lambda_3 w^2 w_3^2
\\ & &
+ w_3 w^2 \sin{\theta} \left( \Re{\bar m}\, \cos{\psi}
+ \Im{\bar m}\, \sin{\psi} \right)
\\ & &
+ w^4 \sin^2{\theta} \left[ \lambda_4
  + \Re{\lambda_5}\, \cos{\left( 2 \psi \right)}
  + \Im{\lambda_5}\, \sin{\left( 2 \psi \right)} \right]
\\ & &
+ w^4 \left( 1 + \cos^2{\theta} \right) \cos{\chi}
\left[ \sin{\left( 2 \psi \right)}\, \Im{\lambda_6}
  + \cos{\left( 2 \psi \right)}\, \Re{\lambda_6} \right]
\\ & &
+ 2 w^4 \cos{\theta} \sin{\chi}
\left[ \sin{\left( 2 \psi \right)}\, \Re{\lambda_6}
  - \cos{\left( 2 \psi \right)}\, \Im{\lambda_6} \right]
\\ & &
+ \lambda_7 w^4 \sin^2{\theta} \cos{\chi}.
\ea
\es
This is minimized relative to the vacuum phase $\chi$,
producing
\bs
\label{ugiofp}
\ba
V_0 &=&
\mu_1 w^2 + \mu_2 w_3^2
+ \lambda_1 w^4 + \lambda_2 w_3^4 + \lambda_3 w^2 w_3^2
\\ & &
+ w_3 w^2 \sin{\theta} \left( \Re{\bar m}\, \cos{\psi}
+ \Im{\bar m}\, \sin{\psi} \right)
\\ & &
+ w^4 \sin^2{\theta} \left[ \lambda_4
  + \Re{\lambda_5}\, \cos{\left( 2 \psi \right)}
  + \Im{\lambda_5}\, \sin{\left( 2 \psi \right)} \right]
\\ & &
- w^4 \left\{
4 \left| \lambda_6 \right|^2 \cos^2{\theta}
+ \left[ \Re{\lambda_6} \cos{\left( 2 \psi \right)}
  + \Im{\lambda_6} \sin{\left( 2 \psi \right)} \right]^2 \sin^4{\theta}
+ \lambda_7^2 \sin^4{\theta}
\right. \\ & & \left.
+ 2 \lambda_7 \sin^2{\theta} \left( 1 + \cos^2{\theta} \right)
\left[ \Re{\lambda_6} \cos{\left( 2 \psi \right)}
  + \Im{\lambda_6} \sin{\left( 2 \psi \right)} \right]
\right\}^{1/2}.
\hspace*{7mm}
\ea
\es
We require $\left| w_1 \right| = \left| w_2 \right|$,
\textit{i.e.}\ either $\theta = \pi / 2$ or $\theta = 3 \pi / 2$
at the minimum of $V_0$ in equation~\eqref{ugiofp}.
We have examined the function
\be
g \left( \theta \right) = A \sin^2{\theta} + B \sin{\theta}
- \sqrt{\cos^2{\theta} + \left( C^2 + D^2 \right) \sin^4{\theta}
  + 2 C D \sin^2{\theta} \left( 1 + \cos^2{\theta} \right)}
\label{ft}
\ee
for various values of the input parameters $A$,
$B$,
$C$,
and $D$ and we have found that,
for instance when\footnote{We only give explicitly
  a continuous range of the parameters of the potential
  for which the wished-for minimum obtains;
  but,
  of course,
  there is a much vaster range of parameters where the same minimim
  also occurs.}
  $-9 < A < -3$,
$-4 < B < -2$,
$-0.2 < C < -0.1$,
and $0.5 < D < 1.2$ the minimum of $g \left( \theta \right)$
always is at the desired value $\theta = \pi / 2$.
Thus,
there is a non-zero-dimension domain of the parameters of the potential
for which its minimum is the desired one.

\section{Confrontation with the phenomenological data}
\label{sec4}

\subsection{Introduction}

We have tested the four sets of conditions
($\alpha \neq \beta \neq \gamma \neq \alpha$)
\bs
\label{ugigfofdp}
\ba
(a) & & \left( \mnu^{-1} \right)_{\alpha \alpha} = 0
\quad \mbox{and} \quad A_{\beta \gamma} = 1/2,
\label{fkpdodi}
\\*[1mm]
(b) & & A_{\alpha \alpha} = 1
\quad \mbox{and} \quad \left( \mnu^{-1} \right)_{\beta \gamma} = 0,
\\*[1mm]
(c) & & \left( \mnu^{-1} \right)_{\alpha \alpha} = 0
\quad \mbox{and} \quad A_{\beta \beta} = A_{\gamma \gamma},
\\*[1mm]
(d) & & \left( \mnu^{-1} \right)_{\alpha \alpha} = 0
\quad \mbox{and} \quad A_{\beta \beta} = A_{\gamma \gamma}^\ast
\ea
\es
against the phenomenological
data~\cite{deSalas:2017kay,Capozzi:2018ubv,Esteban:2018azc},
both for the three choices of $\alpha$
($e$,
$\mu$,
or $\tau$)
and for the two choices of neutrino mass ordering (NO or IO).
Thus,
we have tested 12 different models and,
for each of them,
two mass orderings.
We have found that,
out of the 24 possibilities,
seven models and mass orderings are viable---in the sense
that we shall explain below---\textit{viz.}\ models~1--5 for NO
and models~6 and~7 for IO,
\textit{cf.}\ the listing~\eqref{allmodels2}.
In this section we study in some detail the predictions of each of those models
for the Dirac phase $\delta$ and for the neutrino mass observables,
\textit{viz.}\ the mass of the lightest neutrino $m_\mathrm{minimum}$
($m_\mathrm{minimum} = m_1$ for NO and $m_\mathrm{minimum} = m_3$ for IO),
the total mass of the light neutrinos
\be
\sum m_\nu \equiv m_1 + m_2 + m_3,
\ee
the mass relevant for neutrinoless double-beta decay
\be
m_{\beta\beta} \equiv \left| m_1 c_{12}^2 c_{13}^2
+ m_2 s_{12}^2 c_{13}^2 e^{i \alpha_{21}}
+ m_3 s_{13}^2 e^{i \left( \alpha_{31} - 2 \delta \right)} \right|,
\ee
and the mass relevant for standard $\beta$ decay
\be
m_\mathrm{tritium} = \sqrt{m_1^2 c_{12}^2 c_{13}^2
+ m_2^2 s_{12}^2 c_{13}^2 + m_3^2 s_{13}^2}.
\ee
We recall the cosmological bound~\cite{Akrami:2018vks}
\be
\label{cosmological}
\sum m_\nu < 0.12\,\mathrm{eV},
\ee
which turns out to be relevant in constraining models~4 and~5,
but not the other five models.

We have used as input the nine observables $\delta$,
$\alpha_{21}$,
$\alpha_{31}$,
$s_{12}^2$,
$s_{13}^2$,
$s_{23}^2$,
$m_\mathrm{minimum}$,
$\Delta m^2_\mathrm{solar} \equiv m_2^2 - m_1^2$,
and $\Delta m^2_\mathrm{atmospheric}$.
(Following ref.~\cite{Esteban:2018azc},
we define $\Delta m^2_\mathrm{atmospheric} = m_3^2 - m_1^2 > 0$ for NO
and $\Delta m^2_\mathrm{atmospheric} = m_3^2 - m_2^2 < 0$ for IO.)
For each set of input observables,
we have computed firstly the matrix $\nn$ by using equation~\eqref{diaga2}
and secondly the $A$-matrix elements
$A_{\alpha \beta} = \nn_{\alpha \beta} \left( \nn^{-1} \right)_{\alpha \beta}$.
We have numerically generated thousands of sets
of input observables
that reproduce each of our constraint equations~\eqref{ugigfofdp}
with extremely great accuracy.\footnote{Our method
  differs from the one suggested in a recent paper~\cite{Alcaide:2018vni},
  where the constraint equations are enforced only up to some allowed deviation.
  We use Lagrange multipliers just as ref.~\cite{Alcaide:2018vni} did,
  but in our case their values are much smaller
  than in ref.~\cite{Alcaide:2018vni}
  and therefore the model's constraints are enforced
  to much greater precision.
  Explicitly,
  the constraints $\left( \mnu^{-1} \right)_{\alpha\beta} = 0$ become in our fits
  $\left| \left( \mnu^{-1} \right)_{\alpha\beta} \right|
  \lesssim 10^{-9}\, \mathrm{eV}^{-1}$,
  where `$\lesssim$' stands for ``smaller than and sometimes even
  some orders of magnitude smaller than'';
  the constraints involving the matrix $A$
  are realized with errors $\lesssim 10^{-9}$. \label{foot}}

We have firstly tested our models in the following way.
We have searched for sets of input observables such that
all six observables $s_{12}^2$,
$s_{13}^2$,
$s_{23}^2$,
$\delta$,
$\Delta m^2_\mathrm{solar}$,
and $\Delta m^2_\mathrm{atmospheric}$ are inside their respective
$1\sigma$ Confidence Level (CL) intervals for any one of the three
phenomenological fits~\cite{deSalas:2017kay,Capozzi:2018ubv,Esteban:2018azc}.
If we were able to satisfy
the constraints of one of our models and mass orderings
through observables fully inside the $1\sigma$ ranges of one of the
phenomenological fits,
then we have classified that model and mass ordering as viable.
To be explicit, we have found that both models~1 and~3 for
NO and models~6 and~7 for IO can be met through input observables
inside the $1\sigma$ intervals of either ref.~\cite{deSalas:2017kay},
ref.~\cite{Capozzi:2018ubv}, or ref.~\cite{Esteban:2018azc};
while models~2, 4, and~5 with NO can be satisfied within the
$1\sigma$ domains of both ref.~\cite{deSalas:2017kay}
and ref.~\cite{Capozzi:2018ubv}.
All other models and mass orderings cannot be reproduced
with $1\sigma$~CL input through any of the three phenomenological fits;
therefore we have discarded them.

After this choice of viable models,
we have proceeded to analyze each model in more detail.
We have followed in this endeavour ref.~\cite{Alcaide:2018vni}
and we have used exclusively the phenomenological fit
of ref.~\cite{Esteban:2018azc}.\footnote{We have used the fit
  that does \emph{not}\/ include the Super-Kamiokande atmospheric data,
  \textit{i.e.}\ the fit in the upper part of table~1
  of ref.~\cite{Esteban:2018azc}.}
In ref.~\cite{Esteban:2018azc},
the $\chi^2$ profiles of $s_{23}^2$ and $\delta$
are not symmetrical relative to the best-fit values;
moreover,
those two observables are correlated with each other
much more strongly than (with) the other four oscillation observables.
It makes therefore sense to treat $s_{23}^2$ and $\delta$
differently from the remaining input.

The input values of the observables
never coincide exactly with the best-fit values;
in order to measure the agreement with phenomenology
of each of our `points',
\textit{i.e.}\ sets of input observables,
we have used a function $\chi^2 = \chi^2_{(1)} + \chi^2_{(2)} + \chi^2_{(3)}$.
Here,
\begin{itemize}
\item
  \be
  \chi^2_{(3)} = \left\{ \begin{array}{ll} 0 & \mathrm{for\ NO} \\
    4.71254 & \mathrm{for\ IO} \end{array} \right.
  \ee
  accounts for the fact that the overall quality of the phenomenological fit
  is poorer for IO than for NO.
  The number 4.71254 is the minimum value
  of the quantities $\Delta \chi^2 \left( X \right)$
  (where $X$ is successively $s_{12}^2$,
  $s_{13}^2$,
  $\Delta m^2_\mathrm{solar}$,
  and $\Delta m^2_\mathrm{atmospheric}$)
  depicted in the blue curves of figure~1 of ref.~\cite{Esteban:2018azc}.
  Because of $\chi^2_{(3)}$,
  most fits with IO are of much worse absolute quality than fits with NO,
  in particular our models~6 and~7 fit the data much worse
  than models~1--5.
\item
  \be
  \label{chi111}
  \chi^2_{(1)} \equiv
  \Delta \chi^2 \left( s_{12}^2 \right)
  + \Delta \chi^2 \left( s_{13}^2 \right)
  + \Delta \chi^2 \left( \Delta m^2_\mathrm{solar} \right)
  + \Delta \chi^2 \left( \Delta m^2_\mathrm{atmospheric} \right)
  - 4 \chi^2_{(3)}
  \ee
  is computed from
  the relevant four panels\footnote{$\Delta \chi^2 \left( s_{12}^2 \right)$
    is displayed in the top-left panel,
    $\Delta \chi^2 \left( s_{13}^2 \right)$
    is in the bottom-left panel,
    $\Delta \chi^2 \left( \Delta m^2_\mathrm{solar} \right)$
    is in the top-right panel,
    and $\Delta \chi^2 \left( \Delta m^2_\mathrm{atmospheric} \right)$
    is shown in the middle-right panel
    of figure~1 of ref.~\cite{Esteban:2018azc}.}
  of figure~1 of ref.~\cite{Esteban:2018azc}.
  In those four panels,
  each $\Delta \chi^2 \left( X \right)$
  has been minimized relative to all the observables except $X$.
  In practice,
  we let $s_{12}^2$,
  $s_{13}^2$,
  $\Delta m^2_\mathrm{solar}$,
  and $\Delta m^2_\mathrm{atmospheric}$ vary in their allowed $3\sigma$ ranges,
  \textit{i.e.}\ we allow $\Delta \chi^2 \left( X \right) \leq 9$
  for each $X$.
\item We have computed
  $\chi^2_{(2)} \equiv \Delta \chi^2 \left( s_{23}^2,\, \delta \right)
  - \chi^2_{(3)}$
  by making a two-dimensional interpolation of the values
  of $\Delta \chi^2 \left( s_{23}^2,\, \delta \right)$ that were
  explicitly given for discrete values of $s_{23}^2$ and $\delta$
  in ref.~\cite{Esteban:2018azc}.
\end{itemize}
For a more efficient sampling of the space of input parameters,
we have used global minimization algorithms
and we have performed the minimization of $\chi^2$ for each input point.
Specifically,
we have minimized $\chi^2_{(1)}$
for various fixed $s^2_{23}$ and $\delta$.
(This minimization,
just as all others,
is performed while keeping the conditions of each respective model
obeyed to an extremely high accuracy,
see footnote~\ref{foot}.)
In this way,
at each point in the $s^2_{23}$--$\delta$ plane
we have the minimum relative to all the other oscillation parameters.
In order to find the exact value of $\chi^2_{\mathrm{minimum}}$,
we have added $\Delta \chi^2 \left( s^2_{23} \right)$
and $\Delta \chi^2 \left( \delta \right)$ to $\chi^2_{(1)}$,
because it is much easier to include one-dimensional interpolations
of $\Delta \chi^2 \left( s^2_{23} \right)$
and of $\Delta \chi^2 \left( \delta \right)$ in a FORTRAN code
than to include a two-dimensional interpolation
of $\Delta \chi^2 \left( s_{23}^2,\, \delta \right)$.
Later,
we have recalculated all the discovered input parameters
by using MATHEMATICA with a two-dimensional interpolation
of $\Delta \chi^2 \left( s_{23}^2,\, \delta \right)$.

\subsection{Models 1--5}
\label{sec4.2}

\begin{figure}[t]
\begin{center}
\includegraphics[width=0.95\textwidth]{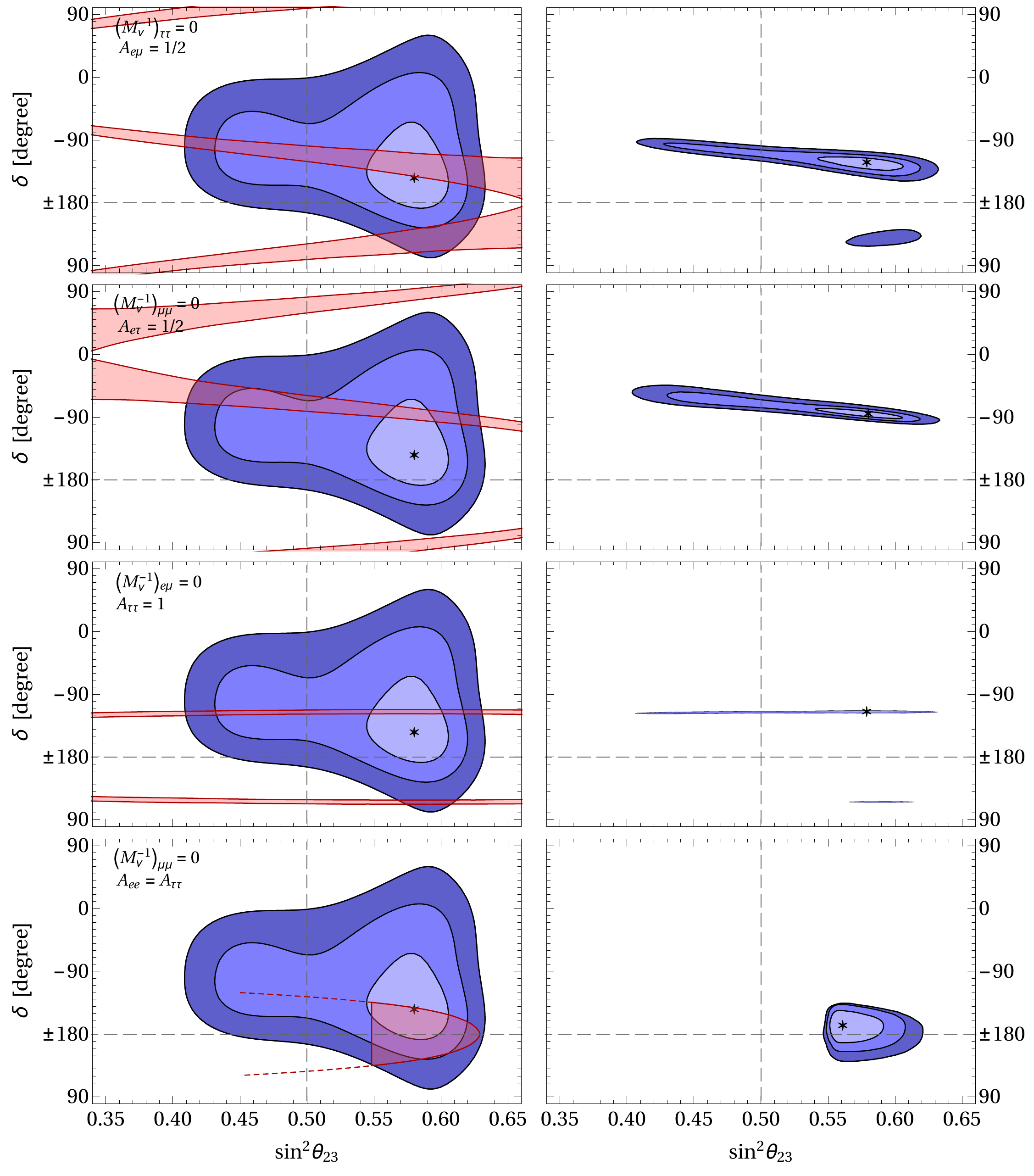}
\vspace{-20pt}
\end{center}
\caption{Left panels: the predictions of each model
  with normal ordering (NO) of the neutrino masses,
  and the phenomenologically allowed areas for NO,
  are displayed in pink and blue colours,
  respectively.
  In the fourth row,
  the conditions of model~4 are obeyed in the whole area
  surrounded by the red line (which might be further extended
  to lower values of $\sin^2{\theta_{23}}$),
  but $\sum m_\nu$ obeys the cosmological bound~\eqref{cosmological}
  only in the pink area.
  Right panels: in different shades of blue,
  the $1\sigma$,
  $2\sigma$,
  and $3\sigma$ regions defined by the simultaneous compliance
  with the model conditions and the phenomenological data.
  In the fourth row,
  only the region satisfying the cosmological bound has been depicted.
  The stars mark the best-fit points.
  Dashed lines at $\delta = \pi$ and $s_{23}^2 = 1/2$
  have been drawn just for orientation.
  More details are given in subsection~\ref{sec4.2}.} \label{fig1}
\end{figure}

Now look at figure~\ref{fig1}.
Each row of that figure corresponds to the model that is
defined by the conditions that are written in the top-left corner
of the left panel of the row,
\textit{viz.}\ to models 1,
2,
3,
and 4,
respectively.
All these models are for a normal ordering of the neutrino masses,
thus $m_\mathrm{minimum} = m_1$.
In figure~\ref{fig1},
just as in figures~\ref{fig2} and~\ref{fig3},
we do not display any panels corresponding to model~5,
because the predictions of models~4 and~5 are almost identical to each other.

\begin{figure}[t]
\begin{center}
\includegraphics[width=1.0\textwidth]{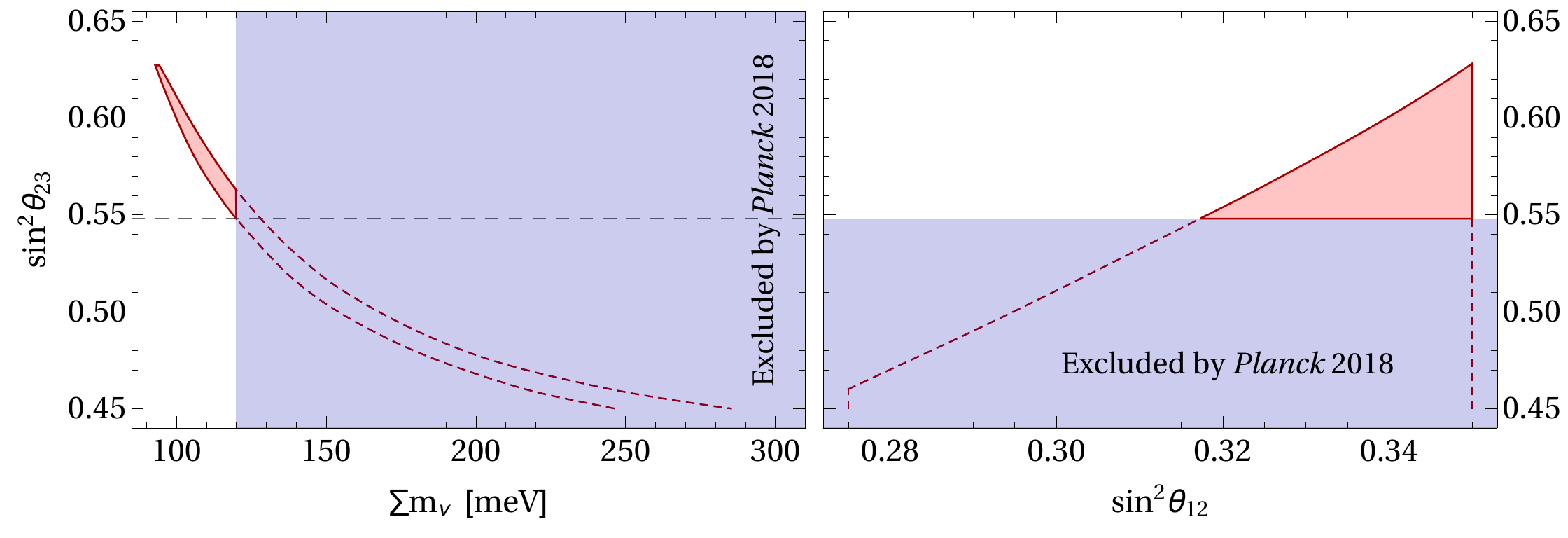}
\vspace{-20pt}
\end{center}
\caption{The correlation among $\sin^2{\theta_{23}}$,
  $\sin^2{\theta_{12}}$,
  and $\sum m_\nu$ in model~4.
  The pink-shaded areas in this figure are equivalent to the pink-shaded area
  in the bottom-left panel of figure~\ref{fig1}.
  A dashed line marks the minimum value 0.548
  of $\sin^2{\theta_{23}}$.} \label{fig8}
\end{figure}

\begin{figure}[t]
\begin{center}
\includegraphics[width=1.0\textwidth]{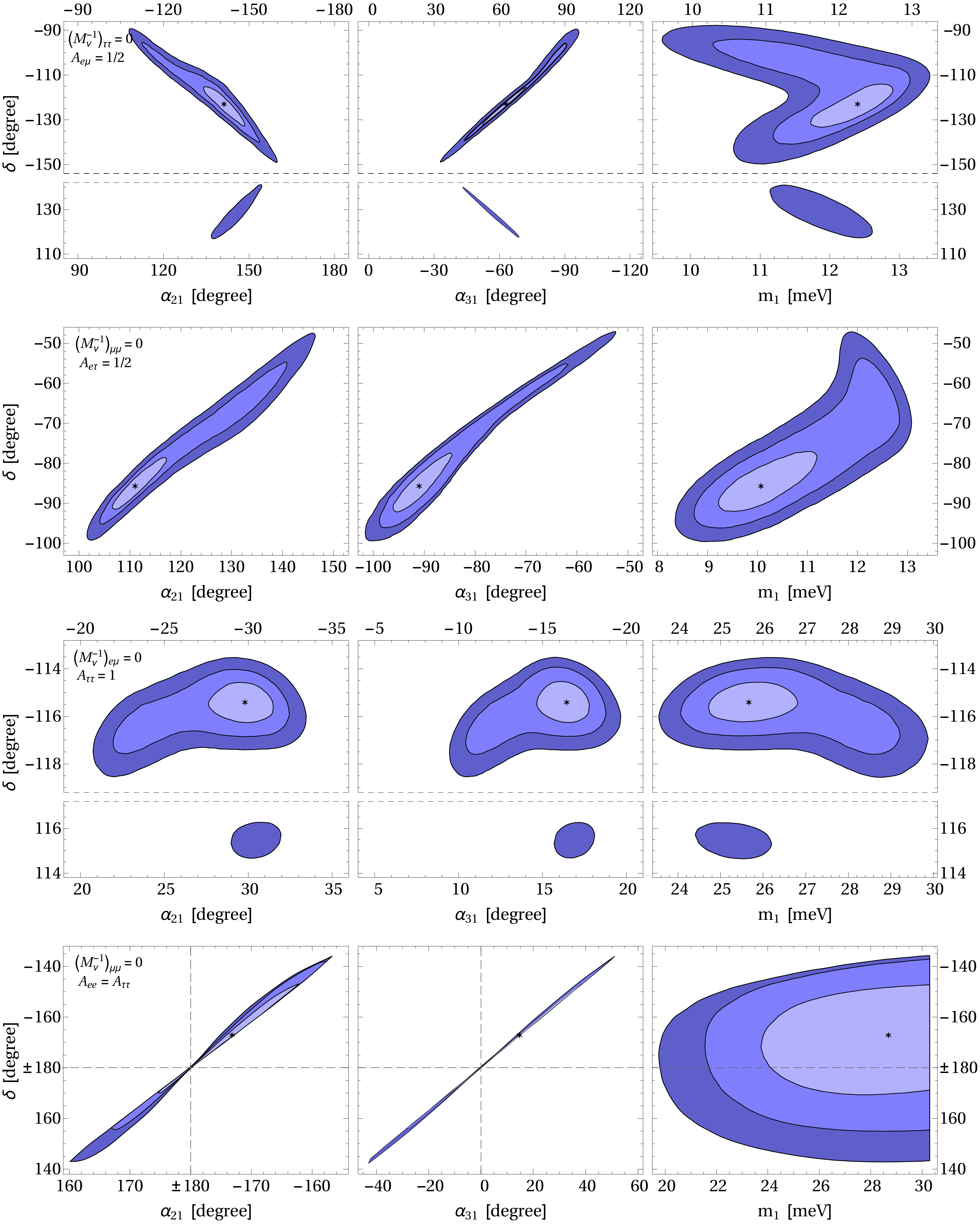}
\vspace{-20pt}
\end{center}
\caption{The same regions as in the right panels of figure~\ref{fig1}
  are now depicted in the $\delta$--$\alpha_{21}$ plane (left panels),
  $\delta$--$\alpha_{31}$ plane (central panels),
  and $\delta$--$m_1$ plane (right panels) for models~1,
  2,
  3,
  and~4,
  respectively,
  from top to bottom.
  For model~4,
  only the points obeying the cosmological bound are displayed.}
  \label{fig2}
\end{figure}

\begin{figure}[t]
\begin{center}
\includegraphics[width=1.0\textwidth]{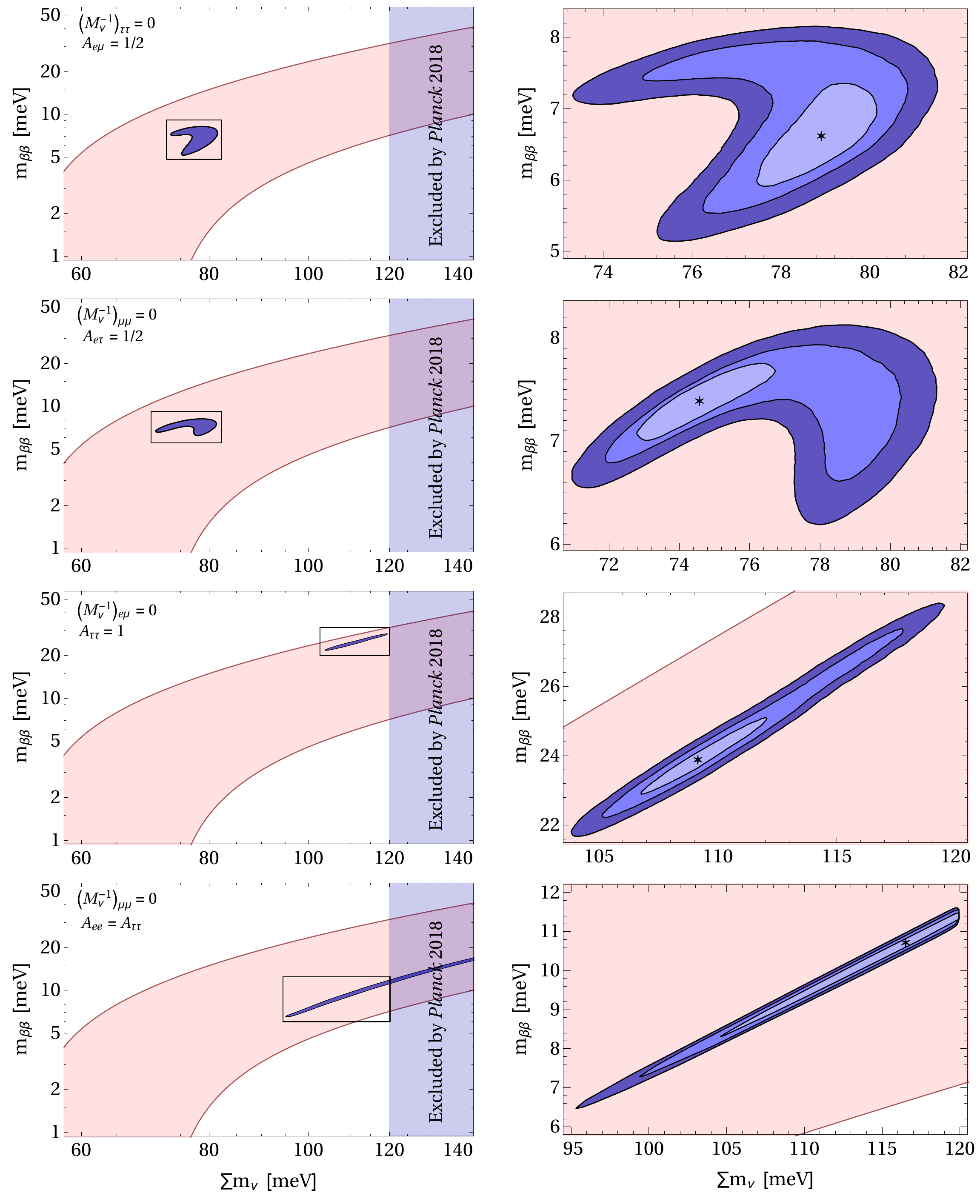}
\vspace{-20pt}
\end{center}
\caption{In the four rows one sees the predictions of models 1, 2, 3, and 4,
  respectively,
  for the sum of the light-neutrino masses
  and for the mass parameter responsible for neutrinoless $2 \beta$ decay.
  Pink areas are allowed by phenomenology alone;
  blue areas include the conditions of each model.
  The right panels are zooms of the marked areas in the left panels.}
\label{fig3}
\end{figure}

In the left panels of figure~\ref{fig1} one sees,
in different shades of blue,
the $1\sigma$,
$2\sigma$,
and $3\sigma$~CL regions in the $s_{23}^2$--$\delta$ plane
that are allowed by the phenomenological data of ref.~\cite{Esteban:2018azc}.
The stars mark the best-fit value of $\left( s_{23}^2,\ \delta \right)$.
The blue regions in the left panels are
identical in all four rows of figure~\ref{fig1}.
The red regions in those panels are specific to each model;
they consist of points that
\begin{description}
\item (a) perfectly obey each model's constraints,
\item (b) satisfy the cosmological bound~\eqref{cosmological},
\item (c) and have $\chi^2_{(1)} - \chi^2_{(1),\mathrm{minimum}} \le 11.83$,
where $\chi^2_{(1),\mathrm{minimum}}$ is the smallest value
of $\chi^2_{(1)}$ in each region of red points;
$\chi^2_{(1)} - \chi^2_{(1),\mathrm{minimum}} \le 11.83$ corresponds
to the $3\sigma$~CL for a Gaussian distribution with two degrees of freedom
(in this case, $s_{23}^2$ and $\delta$).
\end{description}
Comparing the red regions in the top-two left panels of figure~\ref{fig1}
one observes the effects of $\mu$--$\tau$ interchange;
the red bands in the second panel are identical to the ones in the first panel
after the transformation
$s_{23}^2 \to 1 - s_{23}^2,\ \delta \to 180^\circ + \delta$.
In model~4 (the same is valid for model~5)
there is a strong correlation between $s_{23}^2$ and $\sum m_\nu$,
which is depicted in the left panel of figure~\ref{fig8}.
Because of that correlation and of the upper bound~\eqref{cosmological},
$s_{23}^2$ cannot be lower than 0.548,
as depicted through the pink-shaded area in the bottom-left panel
of figure~\ref{fig1}.
If it were not for the bound~\eqref{cosmological},
$s_{23}^2$ would be able to be much lower,
as shown by the dashed red lines in that panel.
Another interesting feature of model~4 is a large forbidden zone
in the $s_{12}^2$--$s_{23}^2$ plane;
that zone,
with low $s_{12}^2$ and high $s_{23}^2$,
can be observed in the right panel of figure~\ref{fig8}.

In the right panels of figure~\ref{fig1} one sees,
for each model~1--4,
the points that have $\chi^2 - \chi^2_\mathrm{minimum}$
smaller than 2.3 ($1\sigma$ or 68.27$\%$ CL),
6.18 ($2\sigma$ or 95.45$\%$ CL),
and 11.83 ($3\sigma$ or 99.73$\%$ CL).
In drawing the right panels we have used the full function
$\chi^2 = \chi^2_{(1)} + \chi^2_{(2)} + \chi^2_{(3)}$
instead of just $\chi^2_{(1)}$ like in the left panels---this
is the reason why the areas in the right panels of figure~\ref{fig1}
are not equal to the intersection of the pink and blue areas in the left
panels;
the pink areas in the left panels were drawn by using
only $\chi^2_{(1)}$ in equation~\eqref{chi111},
while the areas in the right panels were drawn by using
$\chi^2_{(1)} + \Delta \chi^2 \left( s_{23}^2, \delta \right)$.
It should be stressed that,
even though all four right panels of figure~\ref{fig1}
have a light-blue-coloured zone corresponding to
$\chi^2 - \chi^2_\mathrm{minimum} < 2.3$,
that does not mean that all four models~1--4 fit the data equally well,
because $\chi^2_\mathrm{minimum}$ is different for the four models.
The values of $\chi^2_\mathrm{minimum}$ are given in the last row
of table~\ref{table1};
they make clear that models~1 and~3
agree with the data almost perfectly,
while model~2 is not quite as good and model~4
(and also model~5)
is even worse.
For instance,
all the points with $\chi^2 - \chi^2_\mathrm{minimum} < 2.3$ for model~1
have $\chi^2 < 2.7$ and are therefore better than
even the best point of model~4.

One sees in figure~\ref{fig1} that both models~1 and~3
display two different `solutions',
one of them with $\delta \sim -120^\circ$
and the other one with $\delta \approx 120^\circ$.
Under complex conjugation of the lepton mixing matrix,
\textit{i.e.}\ under
$\delta \to - \delta$,
$\alpha_{21} \to - \alpha_{21}$,
and $\alpha_{31} \to - \alpha_{31}$
the conditions defining each model remain invariant,
but the phenomenological bounds on $\delta$ do not;
this is the reason why,
for instance for model~1,
there are two solutions with symmetric values of the phases---but
one of those solutions has much higher values
of $\chi^2 - \chi^2_\mathrm{minimum}$.
For model~3 all the points of the second solution have
$\chi^2 - \chi^2_\mathrm{minimum} > 9$ and therefore that solution
does not appear in table~\ref{table1}.

Comparing the left and right panels of figure~\ref{fig1},
one sees that all models~1--3 severely constrain the phase $\delta$,
but they do not constrain $s_{23}^2$ by themselves alone.
Model~4 has $s_{23}^2$ correlated with $\sum m_\nu$ and,
even when $\sum m_\nu$ becomes very large
(\textit{i.e.}\ when the light neutrinos are almost degenerate),
$s_{23}^2 \gtrsim 0.418$ is constrained;
after the addition of the cosmological bound~\eqref{cosmological}
the constraint becomes much stronger.
Model~4 also restricts $s_{12}^2$,
see the right panel of figure~\ref{fig8}.

Next look at figure~\ref{fig2}.
There,
one sees the same regions as in the right panels of figure~\ref{fig1},
but now displaying the Majorana phases $\alpha_{21}$ and $\alpha_{31}$,
and also the smallest neutrino mass $m_1$,
against $\delta$.
Only points that comply with the cosmological bound on $\sum m_\nu$
are displayed;
this is not an effective constraint for models~1--3,
but it severely constrains model~4 (and model~5).

In figure~\ref{fig3}
one observes the predictions of each model 1--4 for the mass parameters
$\sum m_\nu$ and $m_{\beta \beta}$.
The pink areas in figure~\ref{fig3} are the same for all models
and they are allowed by the phenomenological constraints only;
the areas in various shades of blue are allowed at the $1 \sigma$,
$2 \sigma$,
and $3 \sigma$~CL by the phenomenological constraints
together with each model's conditions.
One sees that each model strongly constrains the mass parameters,
restricting them to a much smaller range
than the one allowed by phenomenology only.
It was already clear from the right panels of figure~\ref{fig2}
that models~1 and~2 work
for much lower values of the neutrino masses than models~3 and~4;
$m_1 \sim 10$~meV for models~1 and~2
while $m_1 \sim 25$~meV for models~3 and~4.
This same fact is observed in figure~\ref{fig3},
where $\sum m_\nu$---but not $m_{\beta \beta}$,
which includes some interference effects---is much higher
in models~3 and~4 than in models~1 and~2.
Notice that a large otherwise-allowed range of model~4
has been eliminated by the cosmological bound on $\sum m_\nu$;
the same may soon happen to model~3,
which predicts $\sum m_\nu \gtrsim 105\,$meV.

We have summarized in table~\ref{table1}
the predictions of each of the models with NO.
\begin{table}
\begin{centering}
\begin{tabular}{||c||c|c||c||c||c||c||}
\hline \hline
model & 1 (1$^\mathrm{st}$ sol.) & 1 (2$^\mathrm{nd}$ sol.) &
2 & 3 & 4 \\
\hline \hline
$m_1 \left( \mbox{meV} \right)$ & 9.9 -- 13.1 & 11.7 -- 12.1 &
8.6 -- 13.0 & 23.9 -- 29.5 & 20.6 -- 30.4
\\ \hline
$\sum m_\nu  \left( \mbox{meV} \right)$ & 74.0 -- 81.2 & 77.7 -- 78.7 &
71.4 -- 80.9 & 104.7 -- 118.6 & 97.2 -- 120.0
\\ \hline
$m_{\beta \beta}  \left( \mbox{meV} \right)$ & 5.4 -- 8.0 & 6.0 -- 6.4 &
6.5 -- 8.0 & 22.0 -- 28.0 & 6.8 -- 11.6
\\ \hline
$m_\mathrm{tritium} \left( \mbox{meV} \right)$ & 13.3 -- 15.9 & 14.7 -- 15.0 &
12.3 -- 15.8 & 25.5 -- 30.8 & 22.5 -- 31.6
\\ \hline \hline
$10 \times s_{23}^2$
& 4.17 -- 6.25 & 5.88 -- 6.03 &
4.17 -- 6.26 & 4.17 -- 6.25 & 5.49 -- 6.14
\\ \hline
$\delta \left( ^\circ \right)$ & 216 -- 268 & 125 -- 132 &
263 -- 309 & 242 -- 246 & 149 -- 224
\\ \hline
$\alpha_{21} \left( ^\circ \right)$ & 204 -- 250 & 143 -- 148 &
103 -- 144 & 327 -- 339 & 163 -- 203
\\ \hline
$\alpha_{31} \left( ^\circ \right)$ & 39 -- 94 & 300 -- 308 &
260 -- 303 & 341 -- 350 & $-37$ -- 51
\\ \hline \hline
$\chi^2_\mathrm{minimum}$ & 0.39 & 8.99 &
1.68 & 0.67 & 3.58
\\ \hline \hline
\end{tabular}
\par\end{centering}
\caption{The $3 \sigma$ bounds for various observables
  in the models with normal neutrino mass ordering.
  These bounds correspond to $\chi^2 - \chi^2_\mathrm{minimum} \le 9$,
  which is equivalent to $3 \sigma$ CL for one degree of freedom;
  they take into account the cosmological bound
  $\sum m_{\nu}<0.12$~eV~\cite{Akrami:2018vks}.
  For model~5 the values are the same as for model~4,
  with the exceptions 
  $10 \times s_{23}^2$ 
  (5.48 to 6.14),
  $\delta$ (154$^\circ$ to 213$^\circ$),
  $\alpha_{31}$ ($-36^\circ$ to $46^\circ$),
  and $\chi^2_\mathrm{minimum}$ (3.82).
  \label{table1} }
\end{table}
In that table we only display points with
$\chi^2 - \chi^2_\mathrm{minimum} \le 9$,
therefore the ranges are somewhat narrower than the ones
observed in the figures,
where the $3\sigma$ regions have $\chi^2 - \chi^2_\mathrm{minimum} \le 11.83$.
For the same reason,
the second solution for model~3 does not appear in table~\ref{table1}.

The observables $s_{12}^2$,
$s_{13}^2$,
$\Delta m^2_\mathrm{solar}$,
and $\Delta m^2_\mathrm{atmospheric}$ are not constrained
by models 1--5,
with the exception $s_{12}^2 \in [0.320,\ 0.350]$ in models 4 and 5;
this is not,
however,
because of the models themselves,
but rather because of the cosmological bound,
that leads those models to necessitate both a rather high $s_{23}^2$
and a rather high $s_{12}^2$.

\subsection{Junction of models~4 and~5}

Models~4 and~5 have almost the same predictions and,
as a matter of fact,
we may join them in only one model,
defined by
\be
\left( \mnu^{-1} \right)_{\mu \mu} = 0
\quad \mbox{and} \quad
A_{ee} = A_{\tau \tau} = A_{\tau \tau}^\ast.
\ee
This model agrees with experiment and has $\chi^2_\mathrm{minimum} = 4.22$,
which is not much worse than either model~4 or model~5 separately.
The CP-violating phases are eliminated:
$\delta = \alpha_{21} = \pi$ and $\alpha_{31} = 0$,
rendering this model CP-conserving in the leptonic sector.
The predictions for the mass observables and for $s_{23}^2$
are exactly the same as the ones displayed in table~\ref{table1} for model~4.

\begin{figure}[t]
\begin{center}
\includegraphics[width=1.0\textwidth]{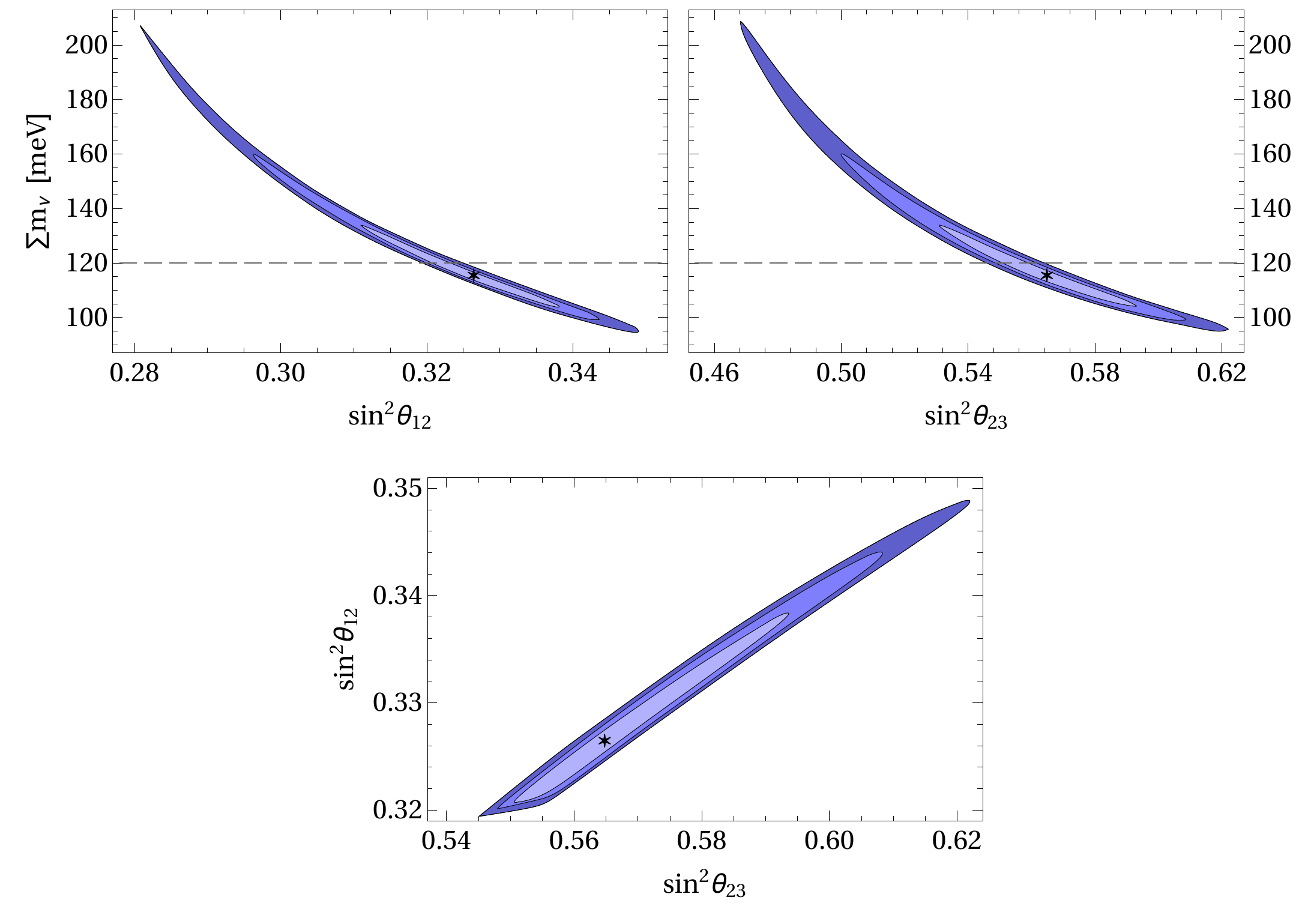}
\vspace{-20pt}
\end{center}
\caption{The correlation among $s_{12}^2$,
  $s_{23}^2$,
  and the cosmological mass in a model uniting models~4 and~5 together.
  In the top row,
  the dashed line represents the cosmological bound~\eqref{cosmological}.
  In the bottom row,
  the depicted areas all respect that bound.}
\label{fig7}
\end{figure}

The plus of this model is that it provides a clear-cut correlation
among $s_{12}^2$,
$s_{23}^2$,
and $\sum m_\nu$.
That correlation is displayed in figure~\ref{fig7}.
On the other hand,
this model requires both $s_{12}^2$ and $s_{23}^2$
to be quite above their best-fit values;
that is the reason why $\chi^2_\mathrm{minimum}$ is rather high for this model.

\subsection{Models~6 and~7}

\begin{figure}[t]
\begin{center}
\includegraphics[width=1.0\textwidth]{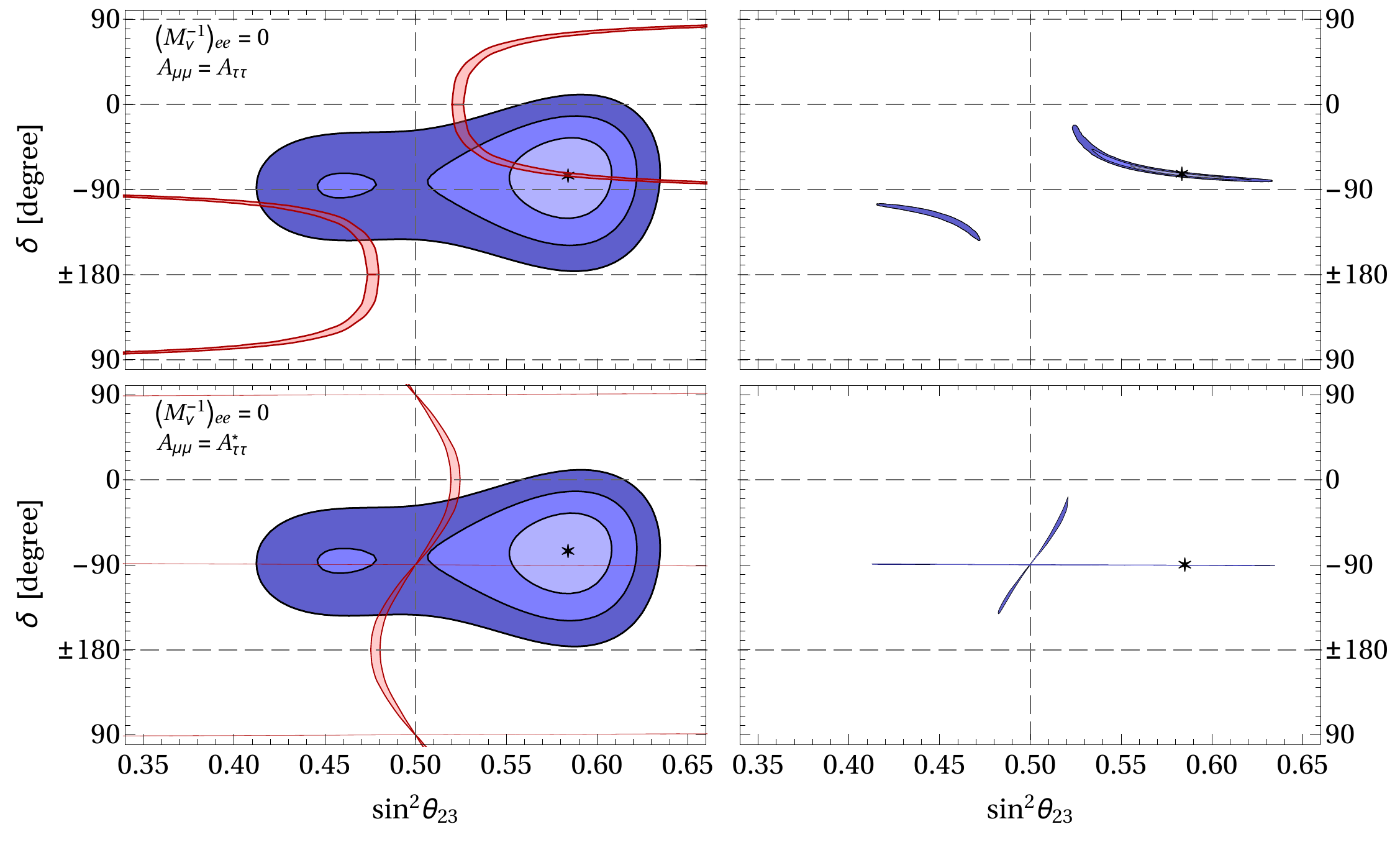}
\vspace{-20pt}
\end{center}
\caption{Left panels: the predictions of each model
  with inverted ordering (IO) of the neutrino masses,
  and the phenomenologically allowed areas for IO,
  are displayed in pink and blue colours,
  respectively.
  The panels in the top row respect model~6
  and the ones in the bottom row are for model~7.
  Right panels: in different shades of blue,
  the $1\sigma$,
  $2\sigma$,
  and $3\sigma$ regions defined by the simultaneous compliance
  with the each model's conditions and the phenomenological data.
  The stars mark the best-fit points.
  Dashed horizontal lines at various values of $\delta$,
  and a vertical dashed line at $s_{23}^2 = 1/2$,
  have been drawn for orientation.
  More details are given in the text.}  \label{fig4}
\end{figure}

\begin{figure}[t]
\begin{center}
\includegraphics[width=1.0\textwidth]{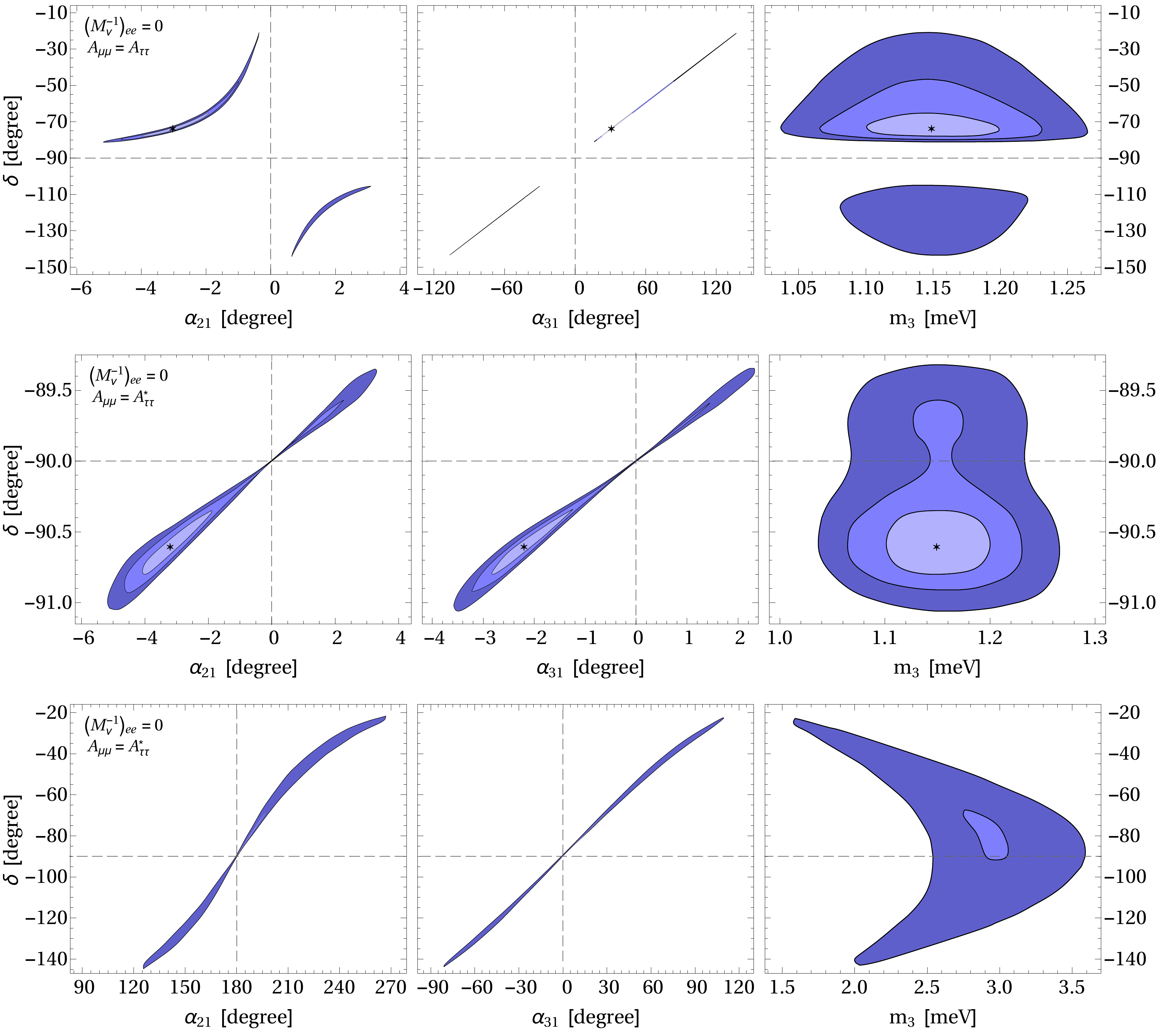}
\vspace{-20pt}
\end{center}
\caption{The same regions as in the right panels of figure~\ref{fig4}
  are now depicted in the $\delta$--$\alpha_{21}$ plane (left panels),
  $\delta$--$\alpha_{31}$ plane (central panels),
  and $\delta$--$m_3$ plane (right panels).
  The top row is for model~6;
  the central and bottom rows are for each of the two solutions of model~7.}
  \label{fig5}
\end{figure}

Figure~\ref{fig4} is analogous to figure~\ref{fig1}.
It features model~6 in its top row and model~7 in its bottom row.
In the pink bands of the left panels
one clearly sees the effect of the $\mu$--$\tau$ interchange symmetry
in the models' defining conditions:
those bands are symmetric under
$s_{23}^2 \to 1 - s_{23}^2,\ \delta \to 180^\circ + \delta$.
In the right panels one sees that both model~6 and model~7
have two different solutions.
Those two solutions are more clearly visible in figure~\ref{fig5},
wherein the first row displays both solutions of model~6 simultaneously
and each of the two lower rows is devoted to one of the solutions of model~7.
One of the two solutions of model~6
has much higher $\chi^2_\mathrm{minimum}$ than the other one.
The two solutions of model~7 are quite distinct,
with the preferred one having
$\alpha_{21} \approx 0$ and $m_\mathrm{minimum} \approx 1.15$ meV,
while the other one has $\alpha_{21} \sim 180^\circ$ 
and $m_\mathrm{minimum}$ twice as large.
Also note,
in the top central panel,
that in model 6 there is an almost perfect linear relation
between $\alpha_{31}$ and $\delta$;
that relation may be expressed by the (approximate) equation
$\alpha_{31} = 178.007^\circ + 1.98254\, \delta$.

\begin{figure}[t]
\begin{center}
\includegraphics[width=1.0\textwidth]{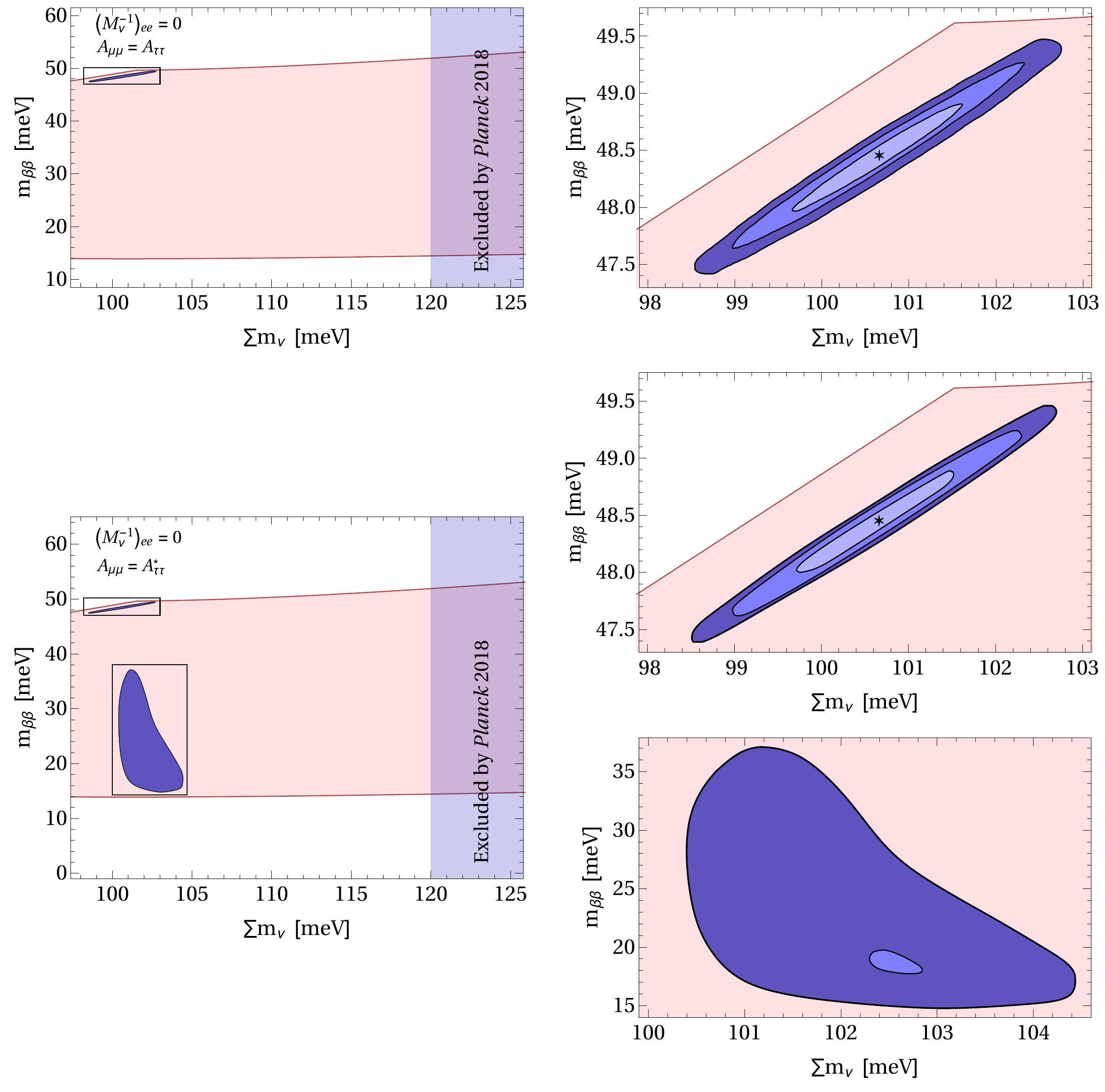}
\vspace{-20pt}
\end{center}
\caption{The predictions of models~6 (top row)
  and~7 (the other two rows) for the sum of the light-neutrino masses
  and for the mass parameter responsible for neutrinoless $2 \beta$ decay.
  The pink areas are the ones allowed by phenomenology alone,
  for an inverted ordering of the neutrino masses;
  the blue areas include the constraints of each model.
  The right panels are zooms of the marked areas in the left panels.}
  \label{fig6}
\end{figure}

In figure~\ref{fig6} one sees the predictions of the two IO models
for the mass parameters.
One observes once again the great difference between the two solutions
of model~7,
with one of them producing a much lower $m_{\beta \beta}$ than the other one.
It is interesting to observe
that both models admit $m_{\beta \beta} \sim 49$ meV,
which is much higher than in the models with NO.

\begin{table}
\begin{centering}
\begin{tabular}{||c||c|c||c|c||}
\hline \hline
model & 6 (1$^\mathrm{st}$ solution) & 6 (2$^\mathrm{nd}$ solution) &
7 (1$^\mathrm{st}$ solution) & 7 (2$^\mathrm{nd}$ solution) \\
\hline \hline
$m_3 \left( \mbox{meV} \right)$ & 1.05 -- 1.25 & 1.11 -- 1.18 &
1.05 -- 1.25 & 1.99 -- 3.39
\\ \hline
$\sum m_\nu  \left( \mbox{meV} \right)$ & 98.7 -- 102.7 & 99.9   -- 101.4 &
98.6 -- 102.6 & 101.2 -- 103.8
\\ \hline
$m_{\beta \beta}  \left( \mbox{meV} \right)$ & 47.5 -- 49.4 & 48.1 -- 48.8 &
47.5 -- 49.4 & 16.0 -- 27.5
\\ \hline
$m_{\mathrm{tritium} }  \left( \mbox{meV} \right)$ & 48.1 -- 50.0 & 48.7 -- 49.4 &
48.1 -- 50.0 & 48.6 -- 49.7
\\ \hline \hline
$10 \times s_{23}^2$
& 5.27 -- 6.27 & 4.29 -- 4.61 &
4.23 -- 6.27 & 4.87 -- 5.18
\\ \hline
$\delta \left( ^\circ \right)$ & 279 -- 326 & 233 -- 251 &
259.0 -- 270.6 & 234 -- 323
\\ \hline
$\alpha_{21} \left( ^\circ \right)$ & 355.2 -- 359.4 & 1.2 -- 2.5 &
$-4.9$ -- 2.9 & 148 -- 233
\\ \hline
$\alpha_{31} \left( ^\circ \right)$ & 17 -- 113 & 287 -- 323 &
$-3.4$ -- 1.9 & $-52$ -- 80
\\ \hline \hline
$\chi^2_\mathrm{minimum}$ & 4.76 & 12.38 &
5.11 & 11.02
\\ \hline \hline
\end{tabular}
\par\end{centering}
\caption{The $3 \sigma$ bounds for various observables
  in the models with inverted neutrino mass ordering.
  These bounds correspond to $\chi^2 - \chi^2_\mathrm{minimum} \le 9$.
  We have included the value $\chi^2_{(3)} = 4.71254$,
  obtained by the NuFIT collaboration,
  in the computation of $\chi^{2}_\mathrm{minimum}$,
  therefore all the $\chi^2_\mathrm{minimum}$ are higher
  than the corresponding values for models with NO in table~\ref{table1}.
  \label{table2} }
\end{table}

\section{Summary and conclusions}
\label{sec5}

In this paper we have shown that four new types of constraints
on the lepton mass matrices,
given in equations~\eqref{ugigfofdp},
can be derived through adequate symmetries imposed on renormalizable models
furnished with three right-handed neutrinos
and a type-I seesaw mechanism.
Each of those constraints leads to
predictive power for the CP-violating phase $\delta$
and for various neutrino-mass quantities.
That predictive power has been studied in some detail
in section~\ref{sec4} of the paper,
especially taking into account the correlations between $\delta$
and the mixing anle $\theta_{23}$ displayed by the phenomenological 
data of ref.~\cite{Esteban:2018azc}.
We have found that a total of seven models are able to fit the data
at the $1\sigma$ level for at least one of the three phenomenological
papers~\cite{deSalas:2017kay,Capozzi:2018ubv,Esteban:2018azc}.
The predictions of each of our models have been given
in tables~\ref{table1} and~\ref{table2}.

\vspace*{5mm}

\paragraph{Acknowledgements:} D.J.\ thanks Jordi Salvado for a detailed
discussion about ref.~\cite{Alcaide:2018vni}.
D.J.\ thanks the Lithuanian Academy of Sciences for support
through projects DaFi2018 and DaFi2019.
The work of L.L.\ is supported by the Portuguese
\textit{Funda\c c\~ao para a Ci\^encia e a Te\-cno\-lo\-gia}\/
through the projects CERN/FIS-PAR/29436/2017,
CERN/FIS-PAR/0004/2017,
and UID/FIS/777/2019;
those projects are partly funded by POCTI (FEDER),
COMPETE,
QREN,
and the European Union.

\end{document}